\documentclass[twocolumn]{aastex701}
\usepackage{appendix}
\newcommand{\ha}{H$\alpha$}

\begin{document}

\title{Inclination Bias in Techniques Used to Identify Be Star Candidates}
\shorttitle{Inclination Bias in Be Stars}
\author[0009-0009-3788-5862, sname=Lailey, gname=Bryan]{B. D. Lailey}
\affiliation{Department of Physics and Astronomy, The University of Western Ontario, 1151 Richmond Street, London N6A 3K7, Canada}
\email[show]{blailey2@uwo.ca}

\author[0000-0002-0803-8615]{T. A. A. Sigut}
\affiliation{Department of Physics and Astronomy, The University of Western Ontario, 1151 Richmond Street, London N6A 3K7, Canada}  
\affiliation{Institute for Earth and Space Exploration (IESX), The University of Western Ontario, Canada}
\email{asigut@uwo.ca}

\correspondingauthor{Bryan Lailey}
\shortauthors{Lailey \& Sigut}

\begin{abstract}
Several methods for identifying Be star candidates are reviewed for observational bias with respect to system inclination, that is the angle between the stellar/disk rotation axis and the observer's line of sight, with focus on two photometric methods that leverage narrow-band filters centred on \ha\ and a spectroscopic method using a \ha\ peak-finding algorithm. Tests for bias were performed using a sample of 20,000 synthetic Be stars drawn from a Salpeter initial mass function and computed libraries of spectral energy distributions and \ha\ profiles. The spectroscopic method showed substantial bias against high inclinations ($i > 80^\circ$). Both photometric methods were biased against low inclinations, with one also biased against inclinations above $80^\circ$, resulting in a surplus in the Be star candidate detection rate for moderate inclinations ($ 50^\circ < i < 80^\circ$). Inclination probability distributions, including the random $\sin i$ factor, are given for the three methods that can be applied to observational samples.  
\end{abstract}

\section{Introduction}

Originally discovered by \cite{Secchi1867}, classical Be (B-emission line) stars are rapidly-rotating, main sequence stars characterized by hydrogen Balmer emission, notably \ha, due to the presence of a circumstellar, decretion disk \citep{Struve1931, Slettebak1982, Porter2003, Rivinius2013}. The nature of the mechanism(s) that puts the disk gas into orbit remains unknown, but is thought to be related to near critical rotation of the central B star \citep{Granada2013}. There are three, mutually inclusive, mechanisms for how this rapid rotation occurs \citep{Navarete2024}: (1) Be stars become rapid rotators via mass transfer from a close binary companion. Evidence from far-ultraviolet spectroscopy \citep{Wang2021} and kinematic modeling \citep{Boubert2018} support this mechanism. (2) Be stars become rapid rotators due to angular momentum transport from core contraction during main-sequence evolution. Here supporting evidence comes from rotating stellar models \citep{Georgy2013} and optical photometry of young open clusters in the Large and Small Magellanic Clouds (LMC, SMC) which found that the fraction of Be stars increases towards the main-sequence turnoff \citep{Milone2018}. (3) Be stars are born rapid rotators. This mechanism is difficult to disentangle from mass transfer from a binary companion because both mechanisms are expected to occur early in a star's lifetime, and an undetected mass transferring-binary might be mistaken for a high intrinsic rotation rate. Nonetheless, studies of the SMC young open cluster NGC~330 have found Be stars at all evolutionary phases suggesting that angular momentum transport cannot be solely responsible for all Be stars' rapid rotation \citep{Iqbal2013,Navarete2024}.

Although Be stars are defined by the presence of $\rm H\alpha$ emission in their spectra, Be star candidates are typically searched for using either low-to-medium resolution spectroscopic or photometric surveys \citep{Rivinius2013}. Ideally, Be star candidates discovered using these less costly methods would then be confirmed using high resolution spectroscopy, although in practice, this may not happen.   

\subsection{The inclination angle and its effect on the search for Be star candidates}
\label{inc}

The inclination angle, $0^\circ\le i\le 90^\circ$, is the angle between a star's rotation axis and the observer's line of sight, with $i=0^\circ$ for pole-on and $i=90^\circ$ for equator-on systems. Both photometric and spectroscopic measurements can be affected by $i$ owing to the presence of the decretion disk in the Be star's equatorial plane, which can add or subtract to the stellar intensity depending on $i$. 

If stellar rotation axes are isotropically oriented in space, the distribution of inclinations for a sample of stars seen by an observer will be $p(i)\,di = \sin{i}\,di$ \citep{gray2021}. This fact forms the basis for statistical tests for correlated stellar spins \citep{Abt2001}. Numerical simulations suggest that the distribution of spin axes in open clusters will be isotropic if small scale turbulence dominates the angular momentum during star formation but will be significantly correlated if the rotational kinetic energy of the cluster dominates during star formation and, moreover, that these correlated spin axes can persist over Gyr timescales \citep{Rey-Raposo2018}. Whether the distribution of inclinations is isotropic, or not, is currently contested. \cite{Corsaro_2017} found evidence of correlated spin axes for the 48 red giant stars with masses between $\rm 1.1-1.7\,M_\odot$ in the old open clusters NGC 6791 and NGC 6819 using asteroseismology. The analysis of \cite{Corsaro_2017} determined that approximately 70 percent of the red giants in the two clusters showed a level of inclination alignment that was extremely unlikely to be observed if the underlying distributions of inclination angles were isotropic, while a sample of 36 field red giant stars was consistent with an isotropic distribution. An analysis by \cite{Gehan2020} supported the conclusion that the inclinations of the field red giant stars are consistent with an isotropic distribution, but the claim of strongly correlated inclinations has been contested by both \cite{Mosser2018} and \cite{Gehan2021} who performed a re-analysis of the spin alignments of both NGC 6791 and NGC 6819 and found no evidence that the inclination distributions deviated from isotropy. 

Be stars offer an alternative to red giant stars to search for evidence of strongly correlated inclinations in open clusters. This is because Be stars are bright, relatively common ($\approx 20\%$ of B-type stars are Be stars \citep{Zorec1997}), and their inclination angles can be reliably determined by three, independent methods: (1) Gravitational-darkening, whereby rapid rotation produces latitudinal variations in stellar intensity \citep{vonZeipel1924} \citep{Townsend2004, Fremat2005,Zorec2016}. (2) Long baseline optical interferometry, whereby the star-disk system of bright and nearby Be stars is directly resolved \citep{vanBelle2012,Sigut2018,Sigut2020}. (3) Spectral synthesis of $\rm H\,\alpha$ morphology, whereby observed $\rm H\,\alpha$ line profiles are compared against a library of synthetic profiles using either figure-of-merit fitting \citep{Sigut2020,Sigut2023} or trained neural networks\footnote{Although machine learning based methods of identifying Be stars are increasingly popular, testing them for inclination bias requires detailed knowledge of both the sample they were trained on and the model parameters which are often not published. Therefore, we will focus on testing classical methods.} \citep{Lailey2023}. Among these methods, (3), the one based on H$\,\alpha$ morphology, is perhaps the simplest from an observational standpoint as it requires only a single observed H$\,\alpha$ line profile of moderate resolution and signal-to-noise.

For Be stars to be a viable probe of correlated spin-axes in open clusters (using any of the above methods), a preliminary question needs to be answered: Are the methods used to find Be star samples biased in any way with respect to inclination? Answering this question is the focus of the present work.  

\subsection{A Lack of High Inclination Be Stars?}
\label{lack_high_i}

Whether the methods used to find Be star samples are biased with respect to inclination can also contribute to the problem of the apparent lack of high inclination Be stars, first noticed by \cite{Rivinius2006} in the data of \cite{Fremat2005}. Using independent methods based on gravitational darkening and $\rm H\,\alpha$ profile fitting, respectively, both \cite{Zorec2016} and \cite{Sigut2023} found that a samples of galactic Be stars contained fewer high inclination stars than expected from the probability distribution of random spin axes, $p(i)\,di = \sin{i}\,di$; the probability of this occurring by chance if the underlying distribution was $\sin i$ is less than $4\,\%$ in both cases. Either the inclinations of the Be star samples are not random (perhaps because the methods used to identify Be star candidates are biased against high inclinations) or both methods used to determine inclinations are biased against high inclinations. As the $\rm H\,\alpha$ profile fitting method used in \cite{Sigut2023} showed no evidence of systematic bias in inclination when tested on a small sample of bright, nearby Be stars with interferometrically resolved disks \citep{Sigut2020}, it is tempting to conclude that the samples really do lack high inclination Be stars. 

Recently, \cite{Cristofari2024} also found evidence of a lack of high inclination stars in the Be star rich, young, open, SMC cluster NGC~330 using high resolution spectroscopy. Although \cite{Cristofari2024} did not calculate inclinations directly, they noted that only two of the 21 Be stars in the cluster had shell parameters (the ratio between the average of the blue and red maxima and $\rm H\,\alpha$ line centre) above 1.5 which suggests that most of the inclinations are below $\rm 75^\circ$ \citep{Hanuschik1996,Sigut2023}. 

\subsection{Sources of Inclination Bias}
\label{malmquist}

Given an observational sample of $N$ stars, one applies a selection criteria based on spectroscopy, photometery, time-series, or some combination, to identify the Be star candidates. If any of the selection criteria use quantities that are correlated with viewing inclination, there is a potential source of bias. In this short section, we look at two obvious examples, the system magnitude (in the V band) and the \ha\ equivalent width (in \AA), defined as
\begin{equation}
\label{eq:ew}
EW \equiv \int \Bigl(\frac{F_\lambda}{F_c}-1\Bigr)\,d\lambda\,,
\end{equation}
where $F_\lambda$ is the flux at wavelength $\lambda$, $F_c$ is the continuum flux, and the integral is over the width of \ha. We adopt the convention that $EW>0$ represents net line emission.  

\begin{figure}
\includegraphics[width=0.48\textwidth ]{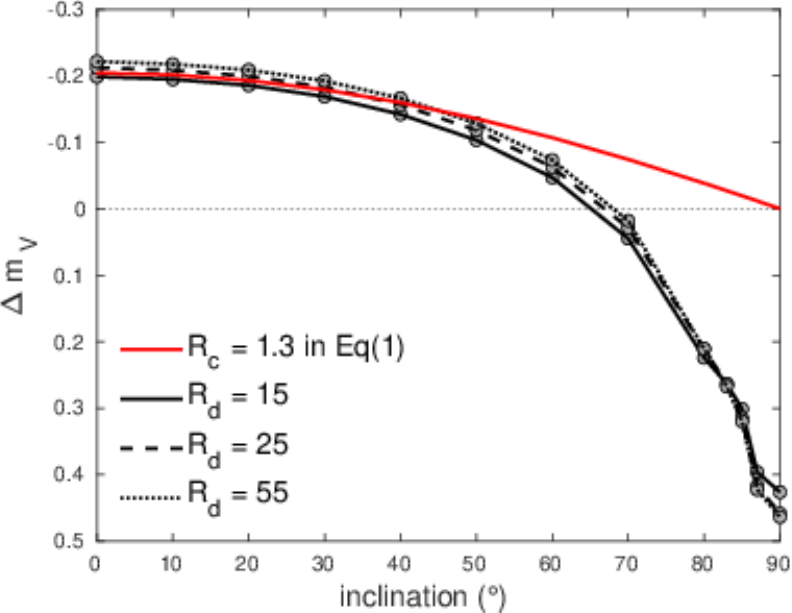}
\caption{Change in the visual magnitude (star+disk minus star alone) of an $M=7\,M_\odot$ Be star as the viewing inclination ranges from $i=0^\circ$ (pole-on star, face-on disk) to $i=90^\circ$ (equator-on star, edge-on disk). The solid lines with symbols are \texttt{Bedisk}/\texttt{Beray} calculations
(see Section~\ref{synthBestars} for computational details). Here the star is surrounded by a disk with parameters $\rho_0=2.35\times 10^{-11}\,\rm g\,cm^{-3}$ and $n=2.25$, and several disk radii are considered (see legend). The red line is the prediction of Equation~\ref{eq:dmv} with $R_c=1.3$ stellar radii.}
\label{fig:VIcorr}
\end{figure}

Figure~\ref{fig:VIcorr} shows the visual V magnitude of a representative Be star as a function of stellar inclination. As V is clearly correlated with inclination, we have the potential conditions for classic Malmquist bias \citep{wall2012}. The basic trend with inclination is easily understood. Assuming both the star and disk radiate as black bodies and the disk's temperature is 60\% that of the star's effective temperature, $T_d=0.6\,T_{\rm eff}$, the magnitude difference between the star+disk system and the star alone is approximately
\begin{equation}
\label{eq:dmv}
\Delta m_V \approx -2.5\log_{10}\left\{1+0.3\,(R_c^2-1)\,\cos i\right\}\;.
\end{equation}
Here the disk is assumed to be optically thick in the continuum (perpendicular through the disk) to radius $R_c$ (measured in stellar radii), and $0.3$ is representative of ratio $B_{\nu}(T_d)/B_{\nu}(T_{\rm eff})$ in the visible over the $T_{\rm eff}$ range of the B stars. The factor of $\cos i$ reflects the reduction in the disk's emitting area with increasing inclination angle. Clearly, this expression is valid only for low-to-moderate inclinations where absorption of starlight by the disk is negligible.  For large inclinations, absorption of starlight by the disk becomes a factor, and the system magnitude eventually returns to the star's magnitude and then further increases due to absorption by the disk, reaching a maximum at $i=90^\circ$. As Be star disks are thin, with $H/R\ll 1$, the system magnitude falling below the stellar magnitude occurs only for high-inclination systems. While the details vary with the general model for the disk density distribution, the overall picture remains the same.

\begin{figure}
\includegraphics[width=0.48\textwidth ]{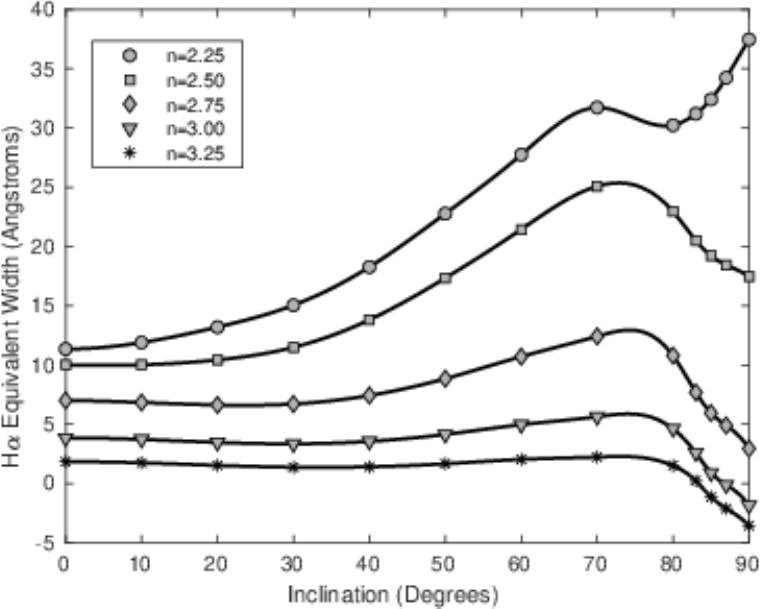}
\caption{Change in H$\alpha$ equivalent width as a function of viewing inclination. Positive equivalent widths represent net emission. The same Be star model of Figure~\ref{fig:VIcorr} was used, except that $R_d$ was fixed at 25 and the power-law index was varied with the values given in the legend.}
\label{fig:ha_inclination}
\end{figure}

Figure~\ref{fig:ha_inclination} shows the H$\alpha$ equivalent width for several representative Be star models (distinguished by the choice of power-law index $n$; see Section~\ref{creating_libraries}) as a function of inclination. Again there is a clear correlation with inclination. The basic trends are easily understood, although there is an interplay of several effects. For high inclinations, $i\ge 80^\circ$, there is a rapid reduction in the EW for most models. Here, the system is sufficiently edge-on that the star is seen through the plane of the (optically-thick) disk, and this leads to so-called shell absorption by the disk material. However, this is not seen for the lowest power-law index model shown, $n=2.25$. Here, the densities in the outer disk are large enough that even viewed edge-on, the disk material not seen against the star contributes substantial emission. Observationally, the H$\alpha$ profile of such objects has large V and R emission peaks separated by a narrow, deep absorption core (zeta\,$\tau$ is a good example). Such Be stars are still classified as shell stars according to the criteria of \cite{Hanuschik1996}, but exhibit large emission equivalent widths. For the other models, the equivalent widths are usually smallest for low-to-moderate inclinations, $i\le30^\circ$, where the low projected disk rotation velocities result in a singly-peaked emission line for H$\alpha$. For increasing inclination, we transition to the classic, doubly-peaked H$\alpha$ line and, therefore, larger equivalent widths. As, nominally, the inclinations follow the $\sin i$ distribution, Be stars with doubly-peaked lines are most common. Finally, for the highest index $n$ models, where the density is low over most of the disk, we see a more-or-less flat behavior until the onset of shell absorption at high inclination.

Finally, it should always be kept in mind that the EW of H$\alpha$ is defined relative to the local continuum which can include a contribution from the disk itself. In most cases, the optical depths in the continuum are quite small; however, for sufficiently dense disks, emission in the adjacent continuum will begin to grow as optical depths build. As the line emission has long since saturated, the H$\alpha$ equivalent width will actually begin to {\it decrease}. Although not shown in Figure~\ref{fig:ha_inclination} for clarity, the EWs for the $n=2.00$ model actually lie below those shown for $n=2.25$ for most inclinations. 

\subsection{Organization}

Section~\ref{sec:methodsid} discusses procedures for identifying Be star candidates using spectroscopy, single exposure photometry, and photometric time-series, using examples drawn from the literature. Section~\ref{synthBestars} describes the computation of the synthetic Be star \ha\ and SED libraries used to test whether methods for identifying Be star candidates are biased in terms of inclination. Section~\ref{sec_spec} tests for inclination bias in spectroscopically determined Be star candidates using the peak-finding algorithm of \cite{Hou2016} on the synthetic Be star spectra. Section~\ref{sec_phot} tests for inclination bias in photometrically determined Be star candidates using the methods of both \cite{Iqbal2013} and \cite{Milone2018}. A discussion of our conclusions follows in Section~\ref{conclusions}.

\section{Methods for Identifying Be Star Candidates}
\label{sec:methodsid}
\subsection{Spectroscopy}
\label{idspec}

Multi-fiber spectroscopic surveys, such as those produced by the Large Sky Area Multi-Object Fiber Spectroscopic Telescope (LAMOST) \citep{Cui2012} and by the Apache Point Observatory Galactic Evolution Experiment (APOGEE) \citep{Majewski2017}, can simultaneously observe entire fields of stars yielding large spectral catalogues. The LAMOST survey contains approximately 11 million spectra with resolution, $\mathcal{R} \equiv \frac{\lambda}{\Delta \lambda}$, $1,800$ and spectral coverage between $\rm 3700-9000\,\AA$, with an additional six million spectra with $\mathcal{R} = 7,500$ and spectral coverage between both $\rm 4900-5400\,\AA$ and $\rm 6000-6800\,\AA$. The APOGEE survey contains approximately 600,000 spectra with $\mathcal{R} = 22,500$ and spectral coverage between $\rm 1.51-1.69 \,\mu m$ \citep{Yan2022}. Contemporary approaches \citep[e.g.,][]{Hou2016,Wang2022} can extract thousands of Be star candidates from these large spectral catalogues often by leveraging machine learning techniques followed by tests to exclude confounding objects such as B[e] and Herbig stars. 

\cite{Hou2016} found 9,752 early-type emission line star candidates while searching dr2 of the LAMOST survey \citep{Cui2012}. A peak-finding routine was used near \ha\ line-centre to identify emission line stars. Once identified, each emission line star was then classified into one of six morphological types by cross-correlation with template line profiles, followed by visual inspection. Any emission line star found within an H\,{\sc ii} region was removed due to the ambiguity of whether the $\rm H \alpha$ emission is intrinsic or extrinsic to the star; Herbig Be stars were separated from classical Be stars using infrared excess measurements from the Two Micron All-Sky Survey (2MASS) \citep{Skrutskie2006} and the Wield-field Infrared Survey Explorer (WISE) \citep{Wright2010}. \cite{Hou2016} culminated in discovering 5,603 new classical Be stars.   

\cite{Shridharan2021} found 3,339 early-type emission line star candidates while searching dr5 of the LAMOST survey \citep{Cui2012}. Again a peak-finding routine was used near \ha\ line-centre to identify emission line stars. Once identified, a series of tests was used to identify confounding objects such as A[e], B[e], and Herbig Be stars in order to remove them. Forbidden emission lines characteristic of A[e] and B[e] stars were identified by fitting the LAMOST spectra to the empirical spectral library of \cite{Blazquez2006} and pre-main-sequence Herbig stars were identified using infrared excess measurements from 2MASS \citep{Skrutskie2006} and WISE \citep{Wright2010}. \cite{Shridharan2021} culminated in identifying 1,089 classical Be stars.   

\cite{Wang2022} found 1,162 Be star candidates by searching dr7 of the LAMOST survey for \ha\ emission in early type stars using the ResNet convolutional neural network \citep{He2016}. The neural network was trained on a combination of Be star spectra (found by searching for local maxima within $\rm 28\,\AA$ of \ha\ line-centre) and non-Be, B-type spectra from the LAMOST survey. Previously discovered Be stars were removed by cross-referencing the literature and confounding Herbig Be stars were removed using infrared excess measurements from the extended Wide-field Infrared Survey Explorer (ALLWISE) \citep{Cutri2021}. \cite{Wang2022} culminated in the discovery of 183 new Be stars.       

\subsection{Photometry}
\label{idphot}

Narrow-band \ha\ filters ($\rm \lambda_{centre} \approx 6563 \,\AA, \Delta\lambda \approx 300\,\AA$) combined with single-exposure optical or infrared photometry allow for the the efficient detection of large numbers of Be star candidates \citep{Rivinius2013}. Methods for identifying Be star candidates using single exposure photometry are particularly well suited to the LMC and SMC where large distances and high cluster densities make spectroscopy difficult \citep{Iqbal2013}. Here, classical, rather than machine learning based, techniques are normally used, with Be star candidates typically determined by comparison to a threshold line in a colour-colour or colour-magnitude diagram.      

\cite{Iqbal2013} found 96 Be star candidates by searching the Optical Gravitational Lensing Experiment (OGLE) catalogue \citep{Paczynski1996} for open clusters in the LMC and SMC with ages between 10 and 100~Myr. Photometric observations using a narrow-band \ha\ filter were compared with observations using Cousins V, R, and I band filters to produce an $\rm R - H\alpha$ vs $\rm V - I$ diagnostic plot. Be star candidates were identified as cluster stars lying above a threshold line in the diagnostic plot. Stars displaying Be characteristics with a V magnitude lower than that of a cluster's main-sequence turnoff were considered to be supergiant B[e] stars and excluded.   

\cite{Milone2018} found that the ratio of Be to early-type main-sequence stars varied between $\approx 0.0$ and 0.6 as a function of the relative magnitude to the main-sequence turnoff by searching Hubble Space Telescope Wide Field Camera 3 (WFC3) \citep{Kimble2008} photometry of 13 young open clusters in the LMC and SMC. Although the focus of this work was to search for evidence of an extended main-sequence consisting of fast rotators, rather than to catalogue Be stars, the method, which involved producing a diagnostic ($\rm F814_{wide}$ vs $\rm F814_{wide}-F656_{narrow}$) colour-magnitude plot where Be star candidates are demarcated from main-sequence B-type stars by a threshold line, is essentially similar to that of \cite{Iqbal2013}.  

\subsection{Photometric Time-Series} 
\label{idtsphot}

Gravitational microlensing surveys such as the MAssive Compact Halo Object project (MACHO) \citep{Alcock2000} and OGLE \citep{Paczynski1996} allowed researchers to identify Be star candidates in the LMC and SMC using a combination of their colours and photometric variability.      

\cite{Mennickent2002} found $\sim$1,000 Be star candidates by searching the OGLE~II catalogue \citep{Zebrun2001} for stars in the SMC with \textit{V}-band absolute magnitudes, \textit{B-V}, and \textit{V-I} colours corresponding to the ranges in which Galactic Be stars are usually found. The \textit{I}-band light curves for each of these $\sim$4,000 stars was then visibly inspected to remove spurious variable stars, Cepheids, and eclipsing binaries, resulting in $\sim$1,000 Be star candidates that were categorized into four types based on their light curve behaviour. The light curves, which were attributed to disk ejection and dissipation, varied on timescales of tens to hundreds of days. The types were assigned based on whether the variation was abrupt or gradual and periodic or stochastic.   

\cite{Vioque2020} found 693 Be star candidates as a side effect of training neural networks to identify Galactic Herbig Be stars using photometric time series data from the Global Astrometric Interferometer for Astrophysics (GAIA) DR2 \citep{Gaia2018}, combined with near and mid-IR data from the 2MASS and WISE surveys, respectively, as well as \ha\ emission data from both the INT/WFC Photometric \ha\ Survey of the Northern Galactic Plane (IPHAS) \citep{Barentsen2014} and the VST Photometric $\rm H\alpha$ Survey of the Southern Galactic Plane (VPHAS+) \citep{Drew2014}.  

\section{Creating the Synthetic Be Stars}
\label{synthBestars}

In order to test methods for identifying Be star candidates, such as those described in the previous section, for inclination bias, 13 large libraries of synthetic Be star SED+\ha\ profiles were generated. Each of the 13 libraries features a central B-type star that corresponds to one of the masses given in Table \ref{stellar_properties}. These stellar masses span most of the mass range of the main sequence B-type stars and are adopted from \cite{Ekstrom2012} assuming a mid-main sequence core hydrogen fraction of $X=0.3$. Within a given library, the synthetic Be stars are differentiated on the basis of their disk density structure and inclination angle as described below.  

\begin{table}
\begin{flushleft}
\begin{tabular}{ccccc} 
\hline\hline
Mass  & Radius  & Luminosity  & $T_{\rm eff}$  & $\log(g)$ \\ [0.49ex] 
($\rm M_{\odot}$) & $\rm (R_{\odot})$ & ($\rm L_{\odot}$) & (K) & \\
\hline
3.00 & 2.9 & $1.10\times 10^2$ & 11,000 & 4.0 \\ 
3.25 & 3.0 & $1.46\times 10^2$ & 11,600 & 4.0 \\ 
3.50 & 3.2 & $2.03\times 10^2$ & 12,200 & 4.0 \\ 
3.75 & 3.3 & $2.62\times 10^2$ & 12,800 & 4.0  \\ 
4.25 & 3.6 & $4.46\times 10^2$ & 14,000 & 4.0\\ 
4.75 & 3.8 & $6.55\times 10^2$ & 15,000 & 4.0\\ 
5.50 & 4.1 & $1.09\times 10^3$ & 16,400 & 4.0\\ 
6.00 & 4.3 & $1.52\times 10^3$ & 17,400 & 3.9 \\ 
8.00 & 5.1 & $4.23\times 10^3$ & 20,600 & 3.9 \\ 
9.00 & 5.4 & $6.28\times 10^3$ & 22,000 & 3.9 \\ 
10.0 & 5.7 & $8.88\times 10^3$ & 23,400 & 3.9 \\ 
12.0 & 6.4 & $1.58\times 10^4$ & 25,600 & 3.9 \\ 
14.0 & 7.0 & $2.51\times 10^4$ & 27,400 & 3.9 \\ 
\hline
\end{tabular}
\end{flushleft}
\caption{Stellar properties adopted from \citet{Ekstrom2012} corresponding to a core hydrogen fraction of $X=0.3$.}
\label{stellar_properties}
\end{table}

\subsection{Creating the libraries of synthetic Be stars}
\label{creating_libraries}

The SED+\ha\ libraries of synthetic Be stars were created using the \texttt{Bedisk} \citep{Sigut2007,Sigut2018} and \texttt{Beray} \citep{Sigut2018} suite of radiative transfer codes. A comprehensive account of using \texttt{Bedisk} and \texttt{Beray} to create libraries of synthetic Be stars can be found in \cite{Sigut2020} and \cite{Sigut2023}. 

The stellar parameters mass, radius, and luminosity $(M_*,R_*,L_*)$ fix the star's $T_{\rm eff}$ and $log(g)$ and, therefore, the star's photoionizing radiation field, which is the principal heating source for the star's equatorial disk. \texttt{Bedisk} is used to compute the temperature structure of the circumstellar disk, given the central star's photoionizing radiation field. 

The circumstellar disk density structure uses a parameterized density model \citep{Sigut2007}. This model depends on $H$, the scale height of the disk, $\rho_0$, the equatorial disk density at the stellar radius, $n$, the power-law index fixing the radial drop-off in disk density, and $r_d$, the disk radius. With these values, the disk density model is
\begin{equation}
\label{eq:diskdensity}
\rho(R,Z) = \rho_0\,\left(\frac{R_*}{R}\right)^n e^{-\left(\frac{Z}{H}\right)^2}\,
\end{equation}
Here, $R$ is the distance from the central B-type star's rotation axis, and $Z$ is the distance above the equatorial plane. Equation \ref{eq:diskdensity} holds for $R_* < R < r_d$ and is zero otherwise. Thus the density structure follows from ($\rho_0$,$n$,$r_d$). The scale height, $H$, of a disk in vertical hydrostatic equilibrium is given by
\begin{equation}
\frac{H}{R} = \frac{c_{\rm s}(T_0)}{V_{\rm K}(R)} \;,  
\end{equation}
where $T_0$, the temperature of the disk in the equatorial plane is set to $60\%$ of the star's $T_{\rm eff}$, $c_s$ is the speed of sound at $T_0$, and ${V_{\rm K}(R)}$ is the Keplerian orbital speed at distance $R$ \citep{Sigut2020}.  

Each of the 13 central B-type star masses listed in Table \ref{stellar_properties} corresponds to a library of synthetic Be stars. Each library has 1,165 different disks, consisting of 15 values of $\rho_0$ evenly distributed in log-space between $10^{-12} \rm \,g\,cm^{-3}$ and $2.5^{-10} \rm \,g\,cm^{-3}$, 11 values of $n$ between 1.5 and 4 in increments of 0.25, and seven values of $r_d$ between 5 $R_*$ and 65 $R_*$ in increments of 10 $R_*$.

The hydrogen level populations computed by \texttt{Bedisk} for each group of 1,155 disks are then used by \texttt{Beray} to compute SEDs and \ha\ profiles for every star-plus-disk system. This is accomplished by solving the radiative transfer equation along a series of rays between the star-plus-disk system and a distant observer \citep{sigut2010,Sigut2018}. Rays that terminate on the surface of the central B-type star use a photospheric boundary condition (either the continuum intensity at at the SED wavelength or a Doppler-shifted, photospheric \ha\ profile); rays that pass through the disk and do not terminate on the central star assume no incident radiation. 

Computing the SEDs and \ha\ profiles introduces the viewing inclination as an additional parameter. A total of 13 different inclination angles ranging from $0^\circ$ to $90^\circ$ in increments of $10^\circ$, plus three additional high inclinations: $83^\circ$, $85^\circ$, and $87^\circ$, are considered. These higher inclinations were included to better trace the development of shell absorption as the central star is increasingly viewed through the plane of the optically-thick disk.

Taking stock, we have 13 mass libraries each with 1,155 disk density models viewed at 13 different inclination angles, totaling 195,195 synthetic Be star SED+\ha\ profiles. Each individual synthetic Be star is represented a \ha\ line profile ($\mathcal{R} = 10,000$) and a visual-IR SED ($\mathcal{R} = 500$). The \ha\ profile, convolved down to $\mathcal{R} = 500$, is also inserted into the SED ensuring that \ha\ is also reflected in any computed colours. 

\subsection{Synthetic Sample of Be stars}
\label{sec_sample}


The first step in the creation of the sample of synthetic Be stars is the random generation of a mass for a Be star following the Salpeter initial mass function \citep{Salpeter1955}. This mass is then used to choose the appropriate SED library from Table \ref{stellar_properties}. Once the mass has been determined, the inclination and the disk parameters described in Equation \ref{eq:diskdensity} are drawn; $i$ and $r_d$ are drawn uniformly from the 13 inclinations and 7 disk radii listed in Section \ref{creating_libraries}. We chose to sample $i$ from a uniform distribution, rather than the observationally expected $\sin{i}$ distribution, because it produces a robust statistical sample across the entire range of inclinations.\footnote{Very low inclinations are expected to be underrepresented using the $\sin{i}$ distribution e.g., the probability of drawing an inclination $i\le \rm 5^\circ$ using the $\sin{i}$ distribution is below $0.4\,\%$ ($\approx 76$ stars) compared to about $7.5\,\%$ ($\approx 1,538$ stars) using the uniform distribution.} For the remaining disk parameters, $\rho_0$ is drawn from a log-normal distribution with $\mu_{\rho_0} = 10^{-11}\rm\,g\,cm^{-3},\,\sigma_{\rho_0} = 5\cdot\,10^{-12}\rm\,g\,cm^{-3}$ and $n$ is drawn from a normal distribution with $\mu_{n} = 2.5,\,\sigma_{n} = 0.5$ \citep[see][]{Sigut2023} before being rounded to the nearest of the 15 disk densities at the stellar radius and 11 power-law indices listed in Section \ref{creating_libraries}. If a synthetic Be star is drawn that is already in the sample, it is discarded. 

To avoid including synthetic Be stars in the sample with disks that are too weak to detect, we compared the \ha\ line profile of the drawn Be star against that of a purely photospheric (diskless) star with the same mass. A figure of merit,
\begin{equation}
\mathcal{F} \equiv \frac{1}{N}\sum_{i=1}^N \bigg| \frac{F_i^{\rm drawn} - F_i^{\rm photo}}{F_i^{\rm drawn}}\bigg|, 
\label{eqn_fom}
\end{equation}
is calculated as the mean absolute percentage difference. If the two spectra differ by more than $2\,\%$, the synthetic Be star is added to the sample; if the two spectra differ by less than $2\,\%$, the synthetic Be star is discarded. Then, this process is repeated until the sample contains 20,000 unique synthetic Be stars.    

\begin{figure}
\includegraphics[width=0.48\textwidth ]{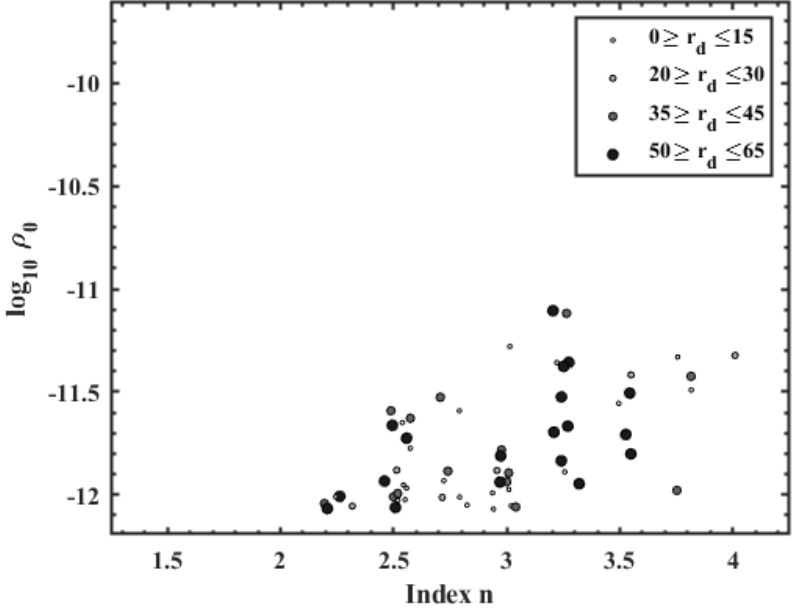}
\caption{A random sample of 1000 discarded disks in the density ($\rho_0)$ versus power-law index ($n$) plane. The disk size ($\rm r_d$) is represented by the symbol size as indicated in the legend.} 
\label{rej_rhovsn}
\end{figure}

The procedure of discarding synthetic Be stars with \ha\ spectra too close to purely photospheric stars of the same mass results in a distribution of some of the synthetic Be star parameters in the sample differing slightly from the distribution they were drawn from. Table \ref{rejdisk_properties} shows the average of the distribution from which the properties were drawn ($\rm \mu_{dist}$), the average of the properties of the discarded Be stars ($\rm \mu_{discard}$), and the average of the properties of the sample ($\rm \mu_{sample}$). Whether a synthetic Be star was discarded for being too similar to a purely photospheric star was only weakly sensitive to the mass of the central B-type star and $r_d$ ($\rm \mu_{sample}$ is within a standard error of $\rm \mu_{dist}$) but depended strongly on $i$, $n$, and $\rho_0$ ($\rm \mu_{sample}$ is several standard errors from $\rm \mu_{dist}$). Weak disks (low $\rho_0$ and high $n$) and larger inclinations were significant contributors to the rejection of a synthetic Be star. Figure \ref{rej_rhovsn} shows the disk parameters of the discarded Be stars on a $\rho_0$ vs $n$ plot where the disk radius is shown by marker size. Note that, for visual clarity, only the disks discarded during the generation of the first 1,000 Be stars are shown. That weak disks are preferentially discarded is shown by the clustering of points in the lower-right section of the plot.   

\begin{figure}
\includegraphics[width=0.48\textwidth ]{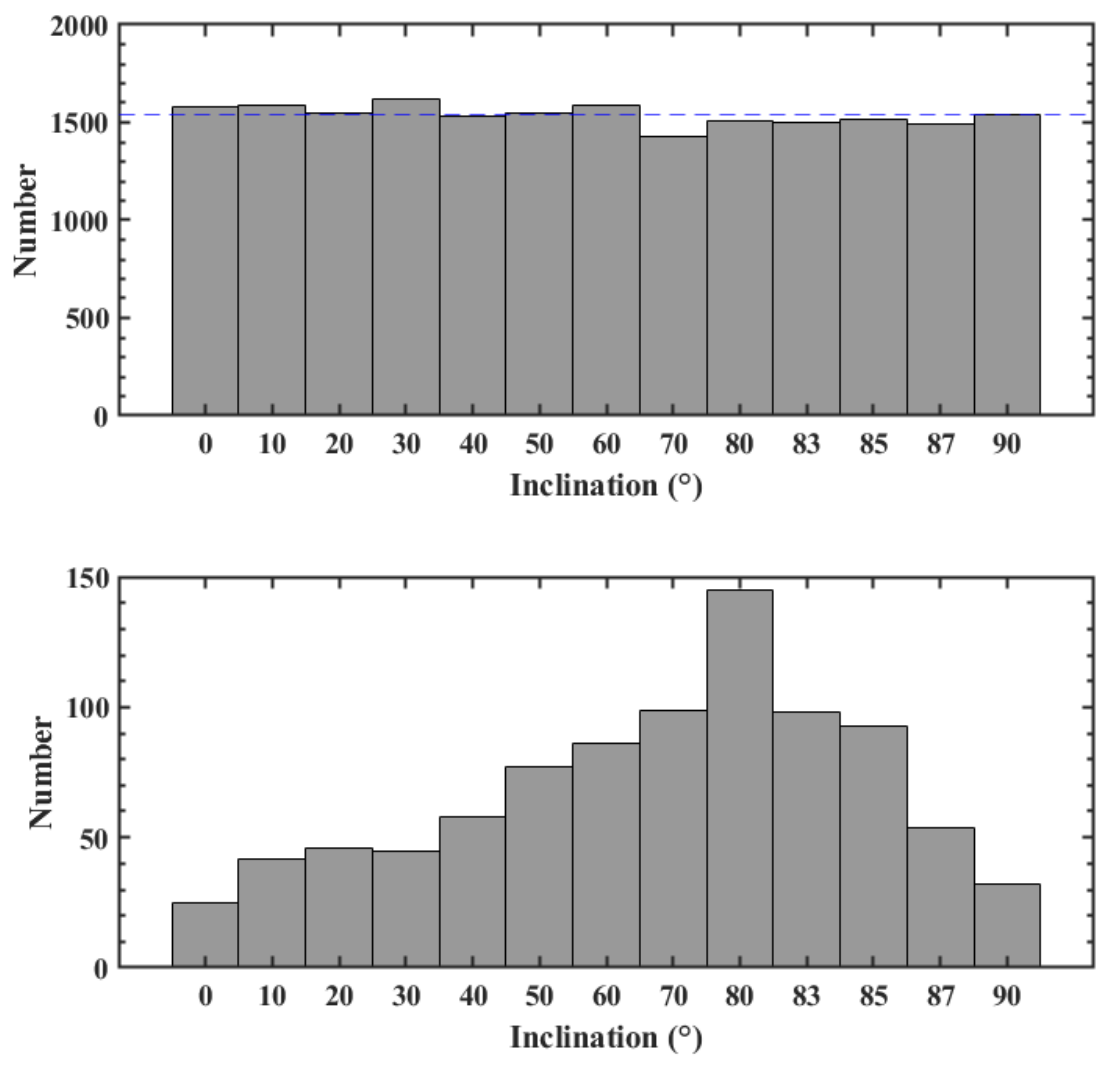}
\caption{Histograms of the number of accepted (top panel) and discarded (bottom panel) synthetic Be stars. The dashed blue line in the top panel shows the expected number if the distribution was perfectly uniform ($\approx1,538$ stars per inclination). Note the large change in the y-axis scale from the top to bottom panel.}
\label{acc_rej_hist}
\end{figure}

Figure \ref{acc_rej_hist} shows a histogram of the number of synthetic Be stars in the sample by inclination (top panel) and another histogram of the number of discarded Be stars by inclination (bottom panel). The bottom panel shows that the likelihood of being discarded due to being too similar to a purely photospheric profile increases with the inclination until $80^\circ$ after which it decreases. This small non-uniformity of the sample is dealt with by giving all of our results as the fraction, rather than the absolute number, of synthetic Be stars rejected by a method to identify candidate Be stars.           
\begin{table}
\begin{flushleft}
\begin{tabular}{lccccc} 
\hline\hline
Parameter & Mass & i & $r_{d}$ & $n$ & $\rho_0$ \\ [0.49ex] 
& ($\rm M_{\odot}$) & $(^\circ)$ & $\rm (R_{\odot})$ & &$\rm (10^{-11} g\,cm^{-3})$    \\
\hline
$\rm \mu_{dist}$ & 4.56 & 54.2 & 35 & 2.5 & 1 \\
$\rm \mu_{\rm discard}$ & 4.53  & $62.9$ & $33.3$ & $3.05$ & $\rm 0.185$ \\
$\overline{\sigma}_{\rm discard}$ &$0.05$  & $0.8$ & $0.5$ & $0.01$ & $\rm0.005$ \\
$\rm \mu_{sample}$ & 4.56  & $53.6$ & $35.1$ & $2.481$ & $\rm 1.07$ \\
$\overline{\sigma}_{\rm sample}$ & $0.01$  & $0.2$ & $0.1$ & $0.003$ & $\rm 0.01$ \\
\hline
\end{tabular}
\end{flushleft}
\caption{Statistical properties of the underlying Be star population (dist), the discarded Be stars with weak disks (discard), and the Be stars retained in the sample (sample). Each parameter (mass, inclination, disk radius, index $n$, and disk density $\rho_0$) is given as a mean $\pm$ standard error ($\mu\pm\overline{\sigma}$), with the exception of the underlying population (dist).}
\label{rejdisk_properties}
\end{table}

\subsection{Synthetic Magnitudes for the Sample}
\label{sec_properties}

Photometric methods for identifying candidate Be stars are based on a threshold line in either a colour-colour or colour-magnitude diagram where the cut is in narrow-band \ha . Photometric magnitudes were computed for several optical filters (denoted by $\beta$) for each synthetic Be star in the sample according to 
\begin{equation}
M_{\beta} = -2.5 \log_{10}\, \int F_{\lambda}\,T^{\beta}_{\lambda}\,d\lambda + K_{\beta}\,, 
\end{equation}
where $T^{\beta}_{\lambda}$ is the normalized transmission curve for the filter $\beta$, $F_{\lambda}$ is the model flux predicted by \texttt{Beray}, and $K_{\beta}$ is a constant to put the colour $\beta$ on the observational system. 

\begin{figure}
\includegraphics[width=0.48\textwidth ]{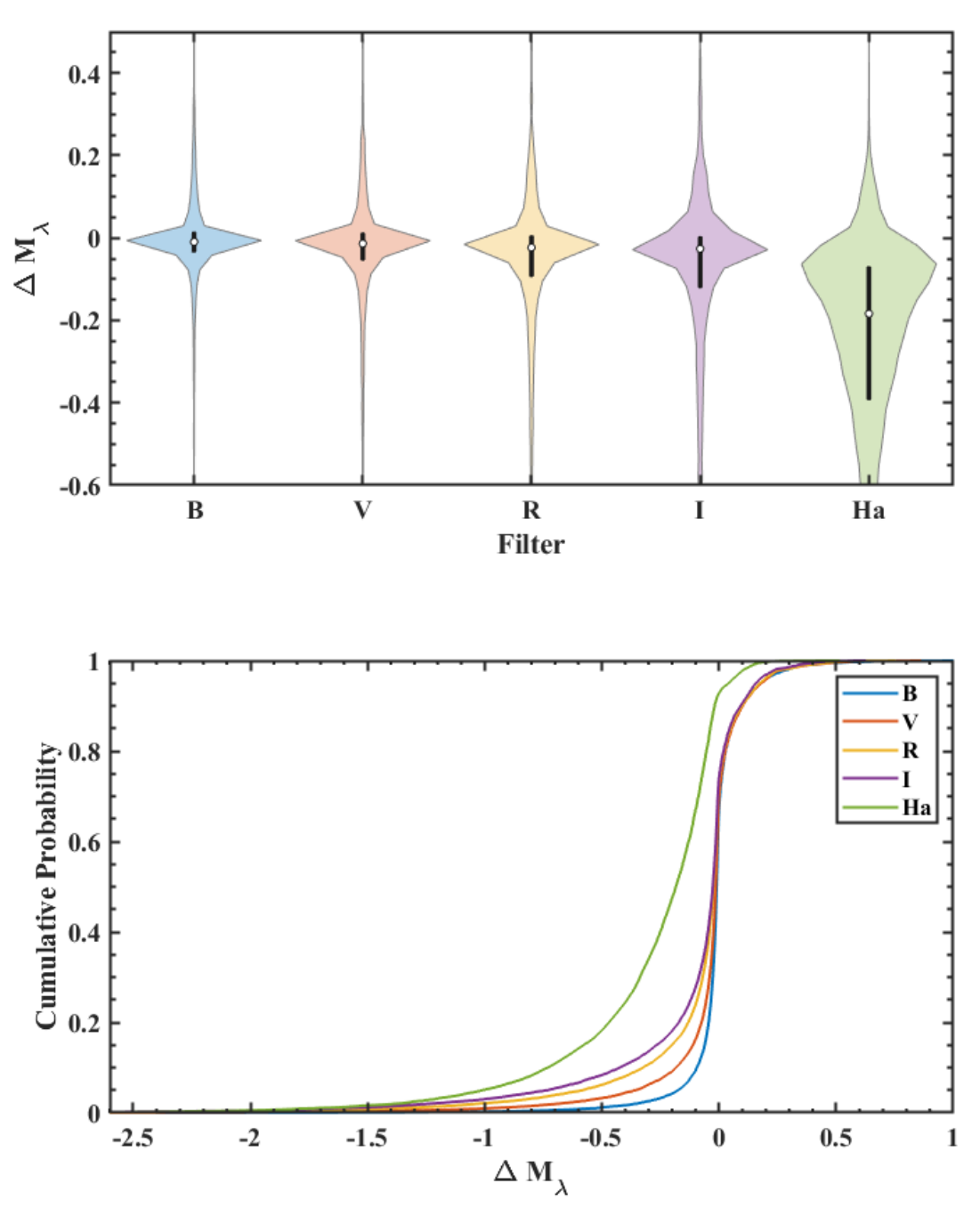}
\caption{{\it Top}: A violin plot of $\Delta\,M_{\beta}$ for the Be star sample in several colours $\beta=\rm BVRI$ and \ha. White circle shows the median of each distribution, and thick black lines show the first and third quartiles, respectively. {\it Bottom}: The corresponding CDFs of the $\Delta\,M_{\beta}$ in the top panel as identified in the legend.} 
\label{violinplot}
\end{figure}

These magnitudes were then compared to a purely photospheric star of the same mass seen through the same filter, $M^*_{\beta}$, to obtain the magnitude difference ($\Delta\,M_{\beta}$ $\equiv  M_{\beta} - M^*_{\beta}$) due to the presence of the circumstellar disk and eliminating the need to determine $K_\beta$. Figure \ref{violinplot} shows the distribution in $\Delta\,M_{\beta}$ between each of the 20,000 Be stars in the sample and a purely photospheric star of the same mass for several optical colours as both a violin plot (top panel) and an empirical cdf (bottom panel). The white circle in the violin plot shows the median of the distribution and the thick black lines show the interquartile range. The significantly increased variability in $\rm H\,\alpha$ magnitudes compared to the other filters is leveraged by most photometric methods to identify candidate Be stars, including the two we evaluate for inclination bias in Section \ref{sec_phot}. Photometric colours (e.g. V-I) are then calculated for each synthetic Be star in the sample and calibrated according to \cite{Ducati2001}.  

\subsection{Sample Spectra and H$\,\alpha$ Equivalent Widths}
\label{sec_haew}

The spectroscopic method of \cite{Hou2016} uses a peak-finding algorithm to search for the presence of $\rm H\,\alpha$ emission in the stellar spectrum. As the method of \cite{Hou2016} was designed to identify Be star candidates using LAMOST spectra, each of the \ha\ profiles in the library of synthetic Be stars has been convolved down to the resolution of LAMOST's low-resolution survey ($\mathcal{R} = 1,800$). A complication resulting from this procedure is that an $\rm H\,\alpha$ profile that differed from a reference photospheric profile of the same mass by at least $2\,\%$ when the library of synthetic Be stars was created (see Equation~\ref{eqn_fom}) may no longer differ by at least $2\,\%$ over the limited wavelength range considered by the method of \cite{Hou2016}. To address this, we re-compared each of the synthetic \ha\ profiles in the 20,000 star library with that of a purely photospheric star of the same mass for the wavelength range used by the method of \cite{Hou2016}. Of the 20,000 profiles, 109 ($\approx 0.5\,\%$) \ha\ profiles differed by less than $2\,\%$ on this reduced wavelength range. These 109 Be stars were discarded from the library for the purposes of determining the inclination bias of the method of \cite{Hou2016}, discussed in Section~\ref{sec_spec}, resulting in a slightly reduced library of 19,891 synthetic Be stars. The full library of 20,000 synthetic Be stars is used for the photometric methods evaluated in Section~\ref{sec_phot}.        

\begin{figure}
\includegraphics[width=0.48\textwidth ]{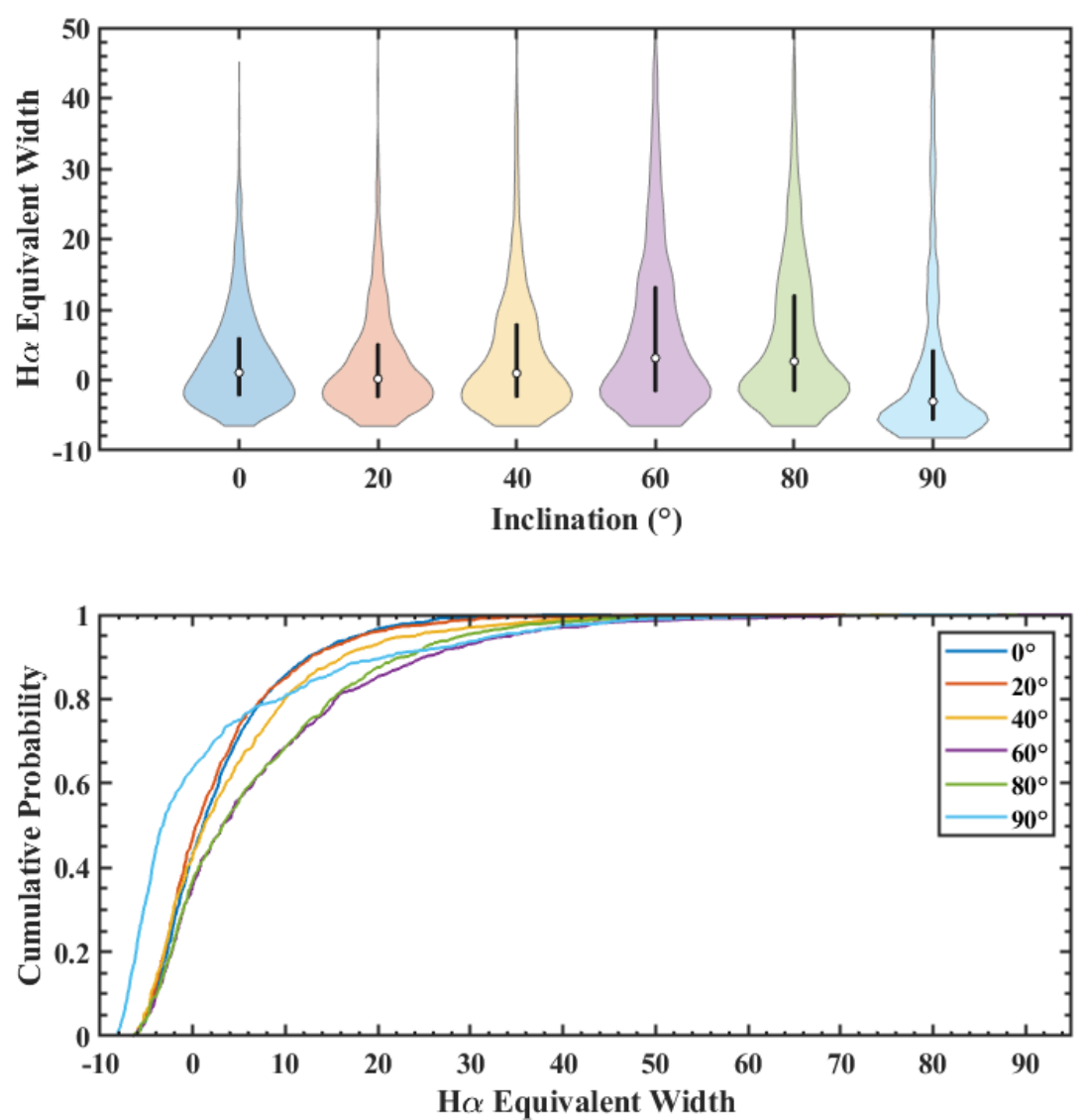}
\caption{{\it Top}: A violin plot of the distribution of $\rm H\,\alpha$ equivalent widths (EW) for several inclinations. White circles show the median of the distribution, and thick black lines show the first and third quartiles, respectively. {\it Bottom}: The corresponding CDFs of $\rm H\,\alpha$ EWs for the same inclinations. Both panels show that they $\rm H\,\alpha$ EWs features a higher fraction of negative values due to shell absorption.} 
\label{violinplotEW}
\end{figure}

Although the method of \cite{Hou2016} is more involved than a simple cut in \ha\ equivalent width, it is still instructive to consider the inclination-dependent variability of the \ha\ equivalent widths. Figure \ref{violinplotEW} shows the distribution of \ha\ equivalent widths for all stars in the sample corresponding to a selection of inclinations, ranging from $i=0^\circ$ to $90^\circ$, as both a violin plot (top panel) and an empirical cdf (bottom panel). The white circle shows the median of the distribution and the thick black lines show the interquartile range. The variability increases with inclination until $60^\circ$, where it plateaus until $80^\circ$, before becoming markedly different at $90^\circ$ as the light from the central B-type star becomes increasingly viewed through the circumstellar disk. Although not definitive in itself, the shape of the distribution in Figure \ref{violinplotEW} suggests that we might expect the spectroscopic method for identifying Be star candidates of \cite{Hou2016} to show an ability to discern between purely photospheric stars and Be star candidates that increases with inclination until it plateaus in the mid-to-high inclinations followed by a drastic change in profile morphology between $80^\circ$ and $90^\circ$.       

\begin{figure}
\includegraphics[width=0.48\textwidth ]{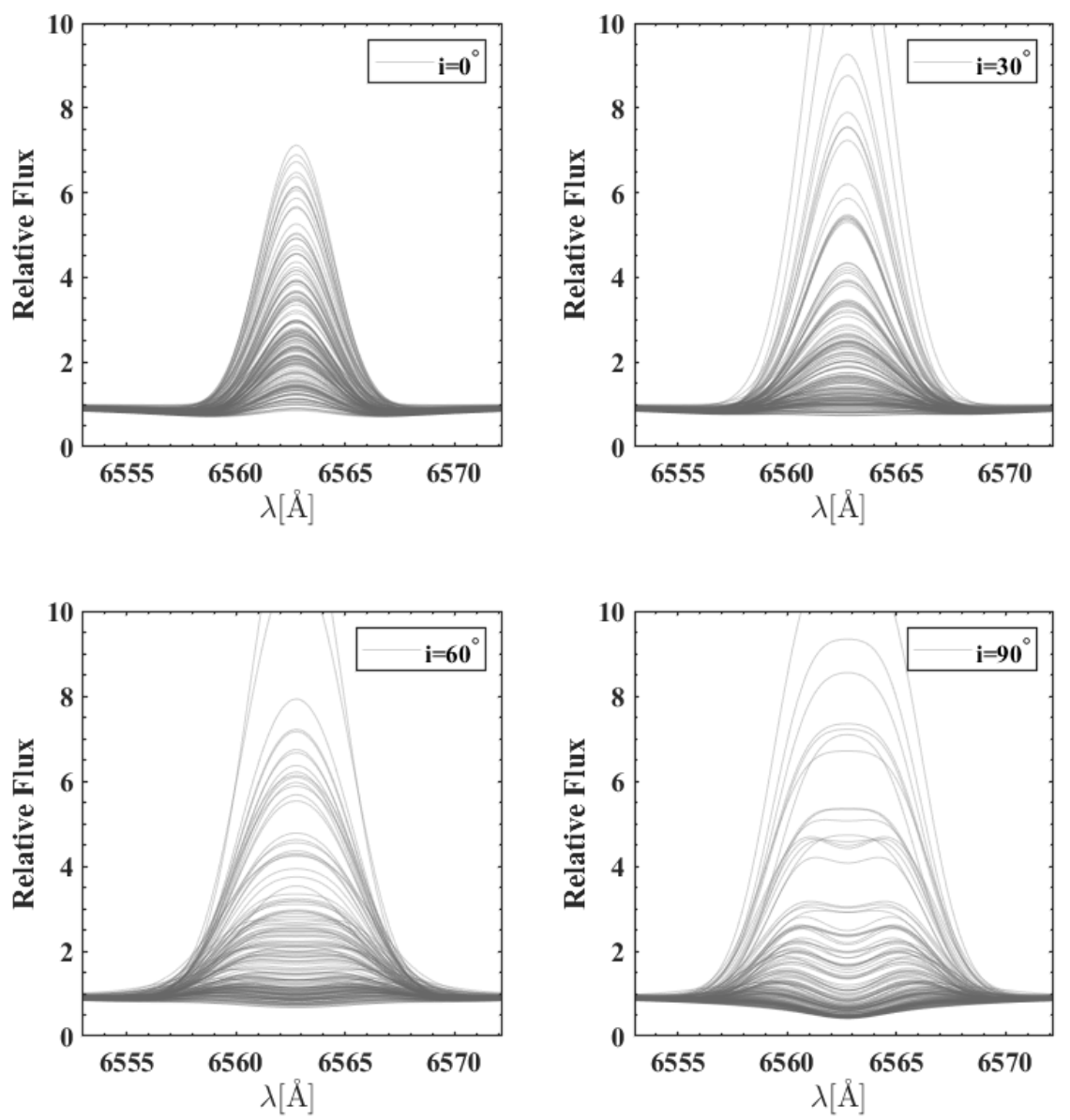}
\caption{Relative flux versus wavelength for sample $\rm H\,\alpha$ profiles convolved down to $\mathcal{R} = 1,800$ at four different inclinations. Each panel shows 150 randomly chosen profiles at that inclination. Note that a few profiles exceed the relative flux limit of 10 on the y-axis.} 
\label{specdistplot}
\end{figure}

%
%
Because the spectroscopic method for identifying Be star candidates of \cite{Hou2016} leverages the morphology of \ha\ profiles, it is instructive to inspect how the distribution of the \ha\ profiles changes with inclination. Figure~\ref{specdistplot} contains relative flux vs wavelength plots for 150 \ha profiles, chosen randomly from the 20,000 star sample at four different inclinations ranging from $0^\circ$ to $90^\circ$. Notable characteristics include the relative lack of variability at $0^\circ$, the dual peaked structure of many $90^\circ$ stars, and shell absorption present at $90^\circ$. As the method of \cite{Hou2016} relies on a peak finding algorithm for identifying Be stars candidates, we can expect significantly more synthetic Be stars to be misidentified at $60^\circ$ and $90^\circ$ due to the lack of a clear peak than at $0^\circ$ and $30^\circ$. At $0^\circ$ none of the plotted Be stars lack a clearly identifiable peak at line centre. 

\section{Bias of a Spectroscopic Method to Identify Be Star Candidates}
\label{sec_spec}

This section will use the method of \cite{Hou2016} to test whether we should expect samples of Be star candidates found using similar spectroscopic methods to be biased in inclination. The method of \cite{Hou2016} is notable because it was designed to identify Be star candidates in bulk by quickly searching through the low-resolution ($\mathcal{R} \approx 1,800$) \ha\ spectra of roughly 200,000 massive stars from dr2 of the LAMOST survey \citep{Cui2012}. 

According to the method of \cite{Hou2016}, Be star candidates are identified using a peak-finding algorithm consisting of two steps that together check three conditions, referred to as $C_1, C_2,$ and $C_3$ below. The first step checks whether the average flux over the 11 pixels centred on \ha\ line centre ($\rm \lambda = 6563\, \AA$) is greater than the flux of the adjacent continuum.
 \begin{equation}
C_1\colon~~~~\frac{1}{11} \sum_{i = -5}^{5} f[n_0 + i]  > f_{\rm continuum}\,,
\end{equation}
where $f[n_0]$ is the flux at line centre, the sum is over pixel index $i$, and $f_{\rm continuum}$ is the flux of the adjacent continuum. Due to the oversampling rate of LAMOST, these 11 pixels correspond to about $16.6\,\AA$. This step is used to find \ha\ lines with emission above the continuum.  

To find weak emission in an absorption trough, the second step checks whether both the average flux of the three pixels centred on \ha\ is greater than the average flux of the five pixels centred on \ha\,
\begin{equation}
C_2\colon~~~~~\frac{1}{3}\sum_{i = -1}^{1} f[n_0 + i]  > \frac{1}{5}\sum_{i = -2}^{2} f[n_0 + i]\,,
\end{equation}
and whether the maximum flux of the five pixels centred on \ha\ occurs within the central three pixels.
\begin{equation}
C_3\colon\max(f[n_0-1:n_0+1]) \geq\max(f[n_0-2:n_0+2])\,,
\end{equation}
where $f[n_0]$ and $i$ are the same as in $C_1$, and the colons in $C_3$ indicate a range over pixel indices. According to the method of \cite{Hou2016}, a Be star candidate is a main sequence, B-type star that satisfies {\em either\/} $C_1$ or both $C_2$ and $C_3$.

A possible complication in evaluating the inclination bias of methods for detecting Be star candidates from LAMOST spectra is the wide range of signal-to-noise ratios (S/N) present in the survey, which range from below $\rm S/N = 10$ to above $\rm S/N = 500$. The analysis below assumes negligible noise, and therefore represents an idealized analysis of the method of \cite{Hou2016}.

\begin{figure}
\includegraphics[width=0.48\textwidth ]{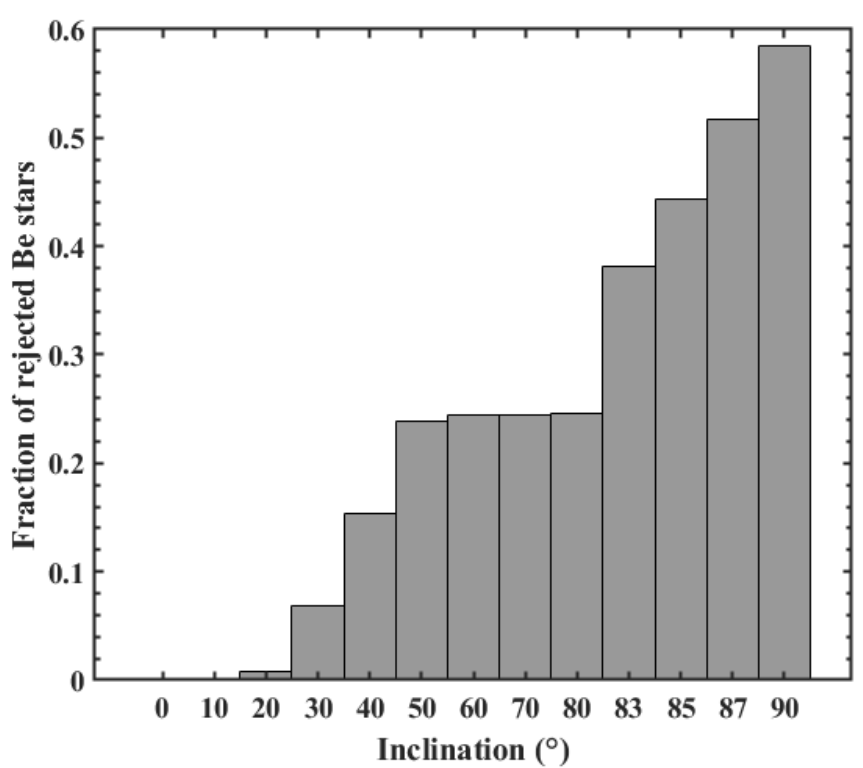}
\caption{The fraction of sample Be stars rejected by the method of \cite{Hou2016} as a function of inclination. This fraction is seen to monotonically increase with inclination. Note the non-linear inclination bins.}   
\label{Hou16plot}
\end{figure}

Figure~\ref{Hou16plot} shows the results of applying the method of \cite{Hou2016} to the sample of synthetic Be stars of Section~\ref{sec_sample} after profiles have been convolved down to the LAMOST resolution of $\rm \mathcal{R} = 1,800$. The vertical axis shows the fraction of Be stars in the sample rejected as Be star candidates, {\it i.e.}, those that don't satisfy either $C_1$ or both $C_2$ and $C_3$, as a function of inclination. Although the sample is constructed such that every star is a Be star (with \ha\ differing from a purely photospheric profile by at least two percent in the mean flux deviation), 4,734 out of 19,891 stars ($\approx 24\%$) are rejected as Be star candidates. No sample Be stars are rejected for $i\le 10^\circ$. However, the rejected fraction rises sharply between $20^\circ$ and $50^\circ$, plateauing until $80^\circ$, and then rises steeply again for $i\ge 80^\circ$. There is clear evidence of a selection bias against very high inclination objects: Be stars with inclinations of $90^\circ$ are more than twice as likely to be rejected as Be stars compared to inclination at or below $80^\circ$. Furthermore, there is clear evidence of a selection bias in favour of low inclination objects, particularly for $i\leq 20^\circ$. 

\subsection{Discussion}
\label{sec_spec_discussion}

The results of Section~\ref{sec_spec} suggest that the method of \cite{Hou2016} for identifying Be star candidates is significantly biased in favour of detecting low inclination angle Be stars and against very high inclination angle Be stars. We now turn to examining the factors that result in this inclination bias.

\begin{figure}
\includegraphics[width=0.48\textwidth ]{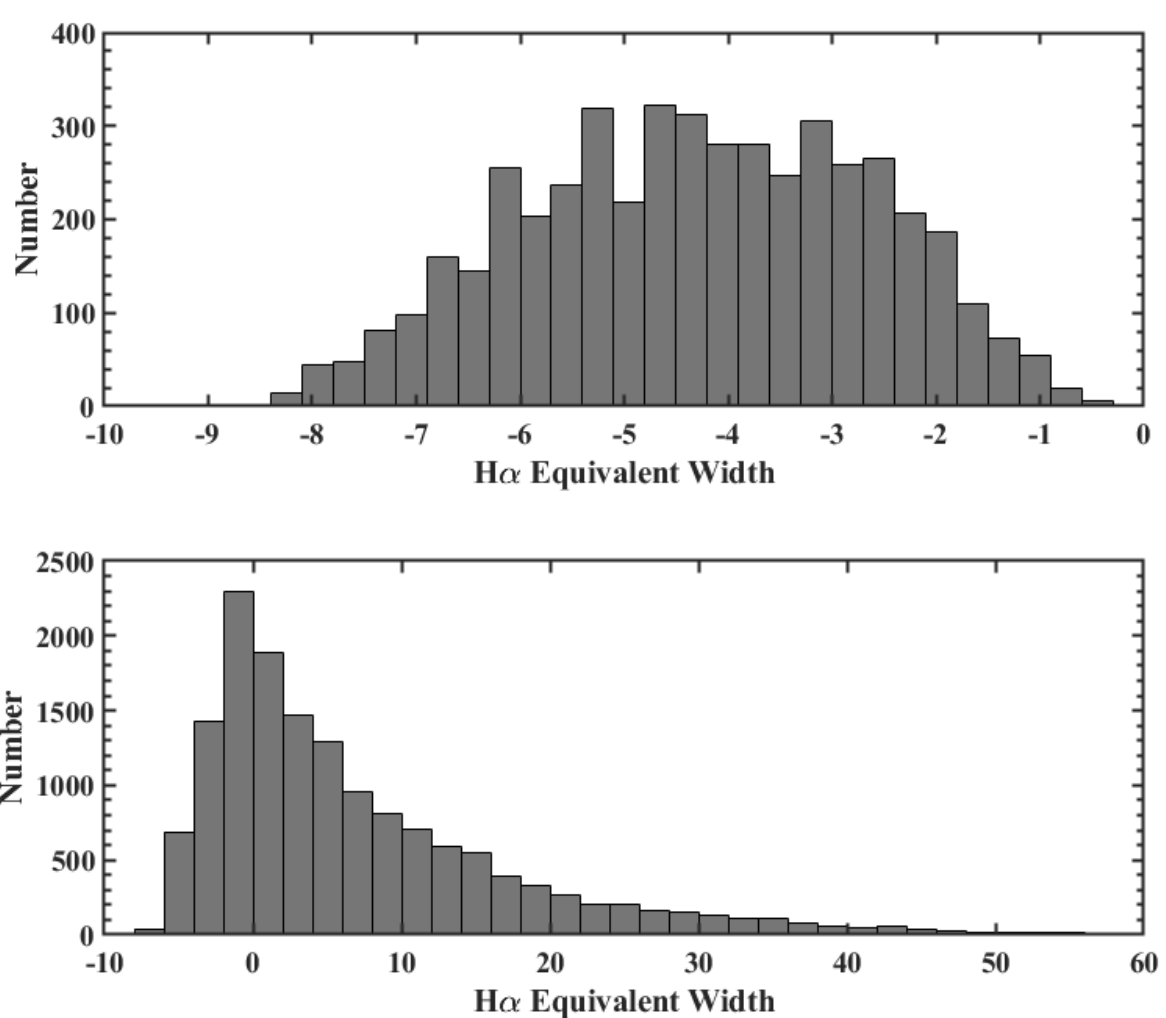}
\caption{{\it Top}: Histogram of the \ha\ EWs of the sample Be stars rejected by the method of \cite{Hou2016}. {\it Bottom}: Histogram of the \ha\ EWs of the sample Be stars accepted by the method of \cite{Hou2016}. Note the change in axis limits from the panel above. There are four sample Be stars with $EW> 60$\AA.}   
\label{hou_haEW_acc_rej_hist}
\end{figure}

Figure~\ref{hou_haEW_acc_rej_hist} shows the distribution of \ha\ equivalent widths for both the synthetic Be stars identified as Be star candidates (bottom panel) and those that were rejected (top panel). The two distributions are very different. Most notably, no Be star with a positive \ha\ equivalent width was rejected as a Be star candidate using the method of \cite{Hou2016} as it would satisfy the requirements of $C_1$. However, while a negative equivalent width was a {\it necessary\/} condition to reject a synthetic Be star as a Be star candidate, Equations $C_2$ and $C_3$ ensured that many of the Be stars with negative \ha\ equivalent widths were correctly identified. 

\begin{figure}
\includegraphics[width=0.48\textwidth ]{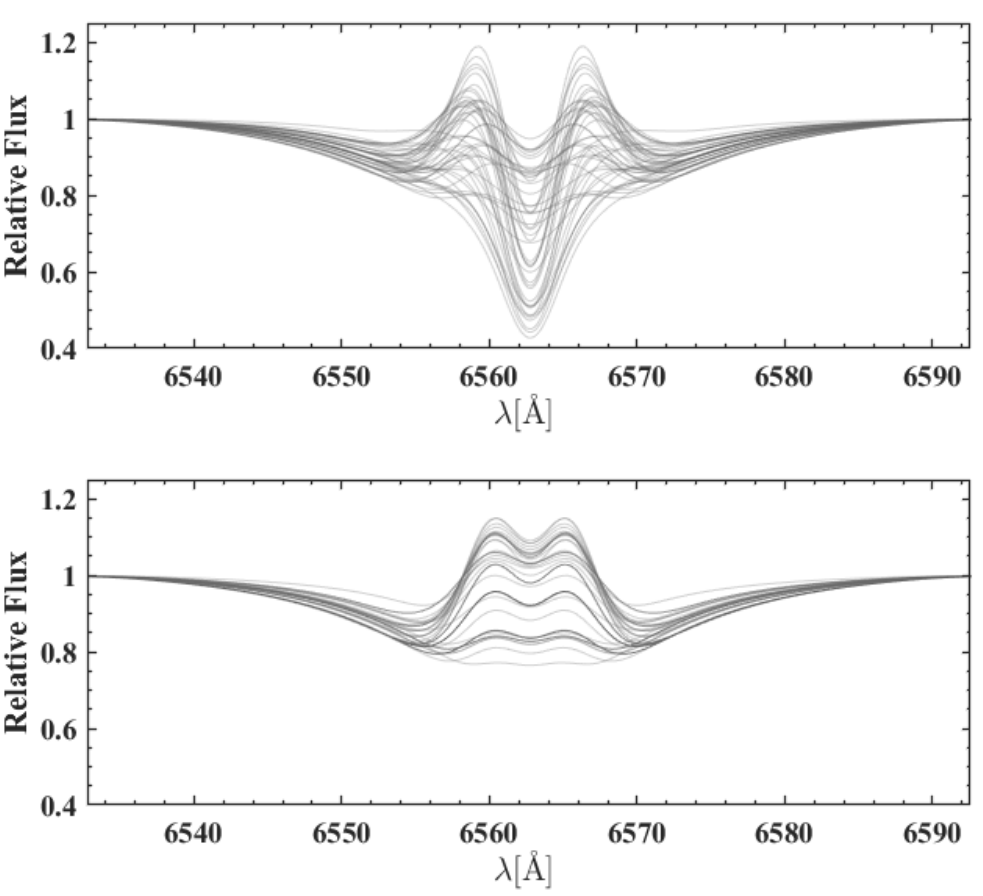}
\caption{{\it Top}: \ha\ line profiles for 50 randomly selected Be stars rejected by the method of \cite{Hou2016} for failing to satisfy each of the conditions $C_1$, $C_2$, and $C_3$. {\it Bottom}: \ha\ line profiles for 50 randomly selected Be star rejected by the method of \cite{Hou2016} for failing to satisfy conditions $C_1$ and $C_3$, {\it but not\/} $C_2$.}   
\label{hou_rej_profiles}
\end{figure}

There are three ways for a synthetic Be star to be rejected as a Be star candidate using the method of \cite{Hou2016}, namely (1) failure to satisfy $C_1$, $C_2$, and $C_3$, (2) failure to satisfy $C_1$ and $C_2$, but not $C_3$, and (3) failure to satisfy $C_1$ and $C_3$, but not $C_2$. In the analysis of the synthetic sample, only (1) and (3) were operative; furthermore, (3) was very rare, with only 66 out of 19,891 ($\approx 0.3\%$) rejected as Be star candidates this way. Turning from the effects of the \ha\ equivalent width to the profiles themselves, Figure~\ref{hou_rej_profiles} is a panel plot of 50 randomly chosen profiles from the synthetic Be star sample that were rejected by the method of \cite{Hou2016} by failing to satisfy $C_1$, $C_2$, and $C_3$ (top panel) and by failing to satisfy $C_1$ and $C_3$, but not $C_2$ (bottom panel). The top panel shows the typical condition for rejection with the \ha\ line featuring more absorption than emission, and there is no visible emission peak inside the absorption trough; this condition is strongly associated with high inclinations and low density disks. The bottom panel shows the rarer condition by which a Be star is rejected because the \ha\ line features more absorption than emission, combined with the separation between the peaks of the doubled-peaked emission structure being too small to satisfy $C_2$ which only occurs at intermediate inclinations $30^\circ \leq i \leq 60^\circ$.      

\begin{figure}
\includegraphics[width=0.48\textwidth ]{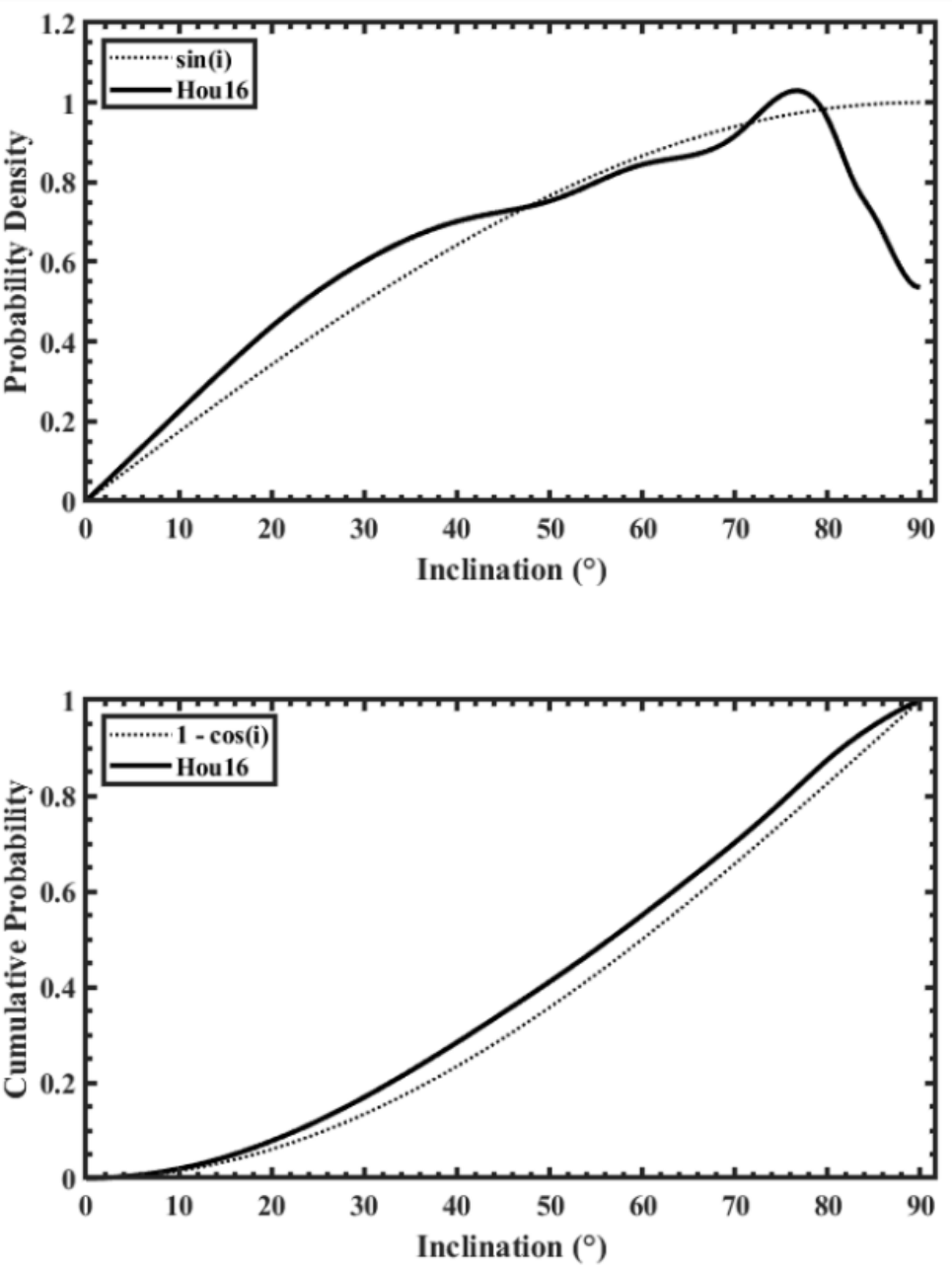}
\caption{The inclination bias of the method of \cite{Hou2016} determined on the sample of synthetic Be star spectra expressed as both a probability density function (top panel) and a cumulative distribution function (bottom panel). In both panels, the inclination bias of Hou16 is shown as a solid line with the $\sin{i}$ probability distribution shown as a dotted line for comparison. Note the strong bias against inclination angles above $80^\circ$.}   
\label{pdf_spectroscopic}
\end{figure}

Finally, Figure~\ref{pdf_spectroscopic} shows the expected observed inclination bias of the method of \cite{Hou2016}, expressed as both a probability density function (pdf, top panel) and as a cumulative distribution function (cdf, bottom panel). This pdf was created to model how an underlying inclination distribution (assumed to be the random $\sin{i}$ distribution) is modified by the bias shown in Figure~\ref{Hou16plot}. We defined an inclination bias function for the method of \cite{Hou2016} on a uniform inclination distribution, $b(i)$, as a cubic spline fit to one minus the fraction of rejected Be stars as a function of inclination shown in Figure~\ref{Hou16plot}. The pdf of the inclination distribution expected for a random, observed sample of Be stars is then
\begin{equation}
\label{eq:ipdf}
 {\rm pdf} =   \frac{b(i)\,\sin\,\!i}{\int_0^\frac{\pi}{2}b(i)\,\sin\,\!i\,di}\;,
\end{equation}
and is shown in the top panel of Figure~\ref{pdf_spectroscopic}. Compared to the $\sin{i}$ distribution, the method of \cite{Hou2016} is modestly biased in favour of low inclinations $i \lesssim 50^\circ$ and heavily biased against very high inclinations $i > 80^\circ$.

\section{Biases of Photometric Methods to Identify Be Star Candidates}
\label{sec_phot}

Although there are methods for identifying Be star candidates using both single exposure photometry and photometric time-series (see Sections~\ref{idphot} \& \ref{idtsphot}, respectively), this work will focus on methods using single exposure photometry because the synthetic Be stars described in Section~\ref{synthBestars} do not have evolving disks. This section will evaluate the methods of \cite{Iqbal2013} and of \cite{Milone2018} to determine if we should expect samples of Be star candidates found using single exposure photometry to be biased in inclination. 

\subsection{Iqbal \& Keller 2013}
\label{sec_ik13}

\cite{Iqbal2013} used observations from the Faulkes Telescope South (FTS) to search for Be stars in young, open clusters in the SMC and LMC. It leverages the fact that stars with strong hydrogen emission will have larger $\rm R - H\alpha$ magnitudes than stars without hydrogen emission at a given $\rm V - I$. According to the method of \cite{Iqbal2013}, a Be star candidate is a main-sequence, B-type star that lies above a threshold line on an $\rm R - H\alpha$ vs $\rm V - I$ colour-colour diagram. One complication that arises is that $\rm R - H\alpha$ has an arbitrary zero point; \cite{Iqbal2013} addressed this issue by calibrating such that main-sequence stars are tightly clustered around $\rm R - H\alpha = 0$. Be star candidates identified by this method represent a lower-bound to the the number of Be stars in a cluster because the method is inherently biased against small emission equivalent widths (in addition, of course, to those missed due to the transient nature of the Be phenomena). 

Recently, \cite{Navarete2024} expanded on the work of \cite{Iqbal2013} using the Southern Astrophysical Research Telescope (SOAR) Adaptative Module Imager (SAMI), which provides high angular resolution, adaptive optics assisted photometry \citep{Tokovinin2016}. Unlike observations from the FTS, which are both seeing and flux limited, SAMI offers full photometric coverage up to the early A-type stars and is able to resolve stars in the densely packed cores of young open SMC clusters \citep{Navarete2024}. Owing to the high angular resolution of SAMI, \cite{Navarete2024} found approximately twice as many Be star candidates as \cite{Iqbal2013} by replicating their method with a very slightly modified threshold line. Preliminary testing performed during the course of this work found that the effect of \cite{Navarete2024}'s modification to the threshold line of \cite{Iqbal2013} on the inclination bias was negligible; thus, the inclination bias determined in this section will also apply to \cite{Navarete2024}. 

\begin{figure}
\includegraphics[width=0.48\textwidth ]{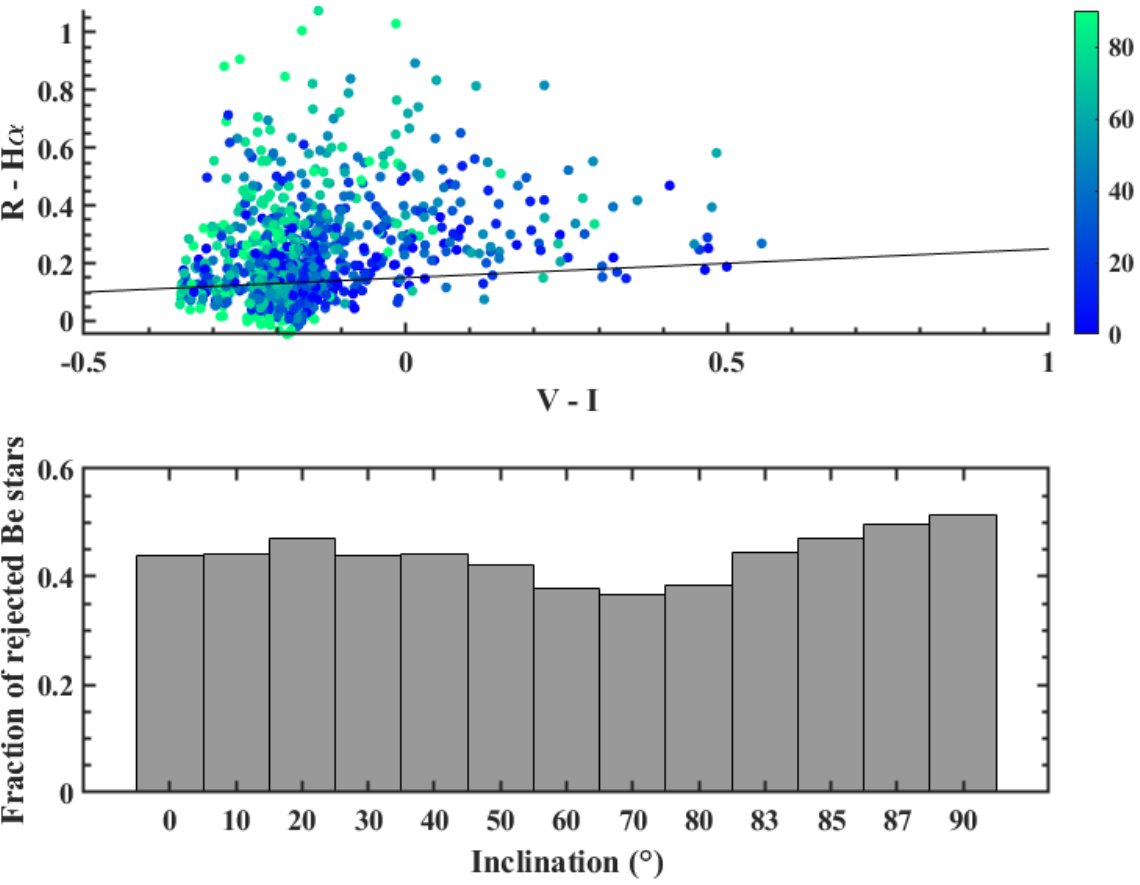}
\caption{{\it Top}: An $(\rm R-H\alpha)$ vs $(\rm V-I)$ colour-colour diagram used to identify Be star candidates based on the method of \cite{Iqbal2013}. Shown as circles are 1,000 randomly selected sample Be stars, with inclination coded according to the colourbar on the right. The black line is the threshold above which stars are considered Be star candidates. {\it Bottom}: A histogram of the fraction of all sample Be stars that were rejected as Be star candidates as a function of inclination.} 
\label{ikplot}
\end{figure}

The top panel of Figure~\ref{ikplot} shows the method of \cite{Iqbal2013} applied to the synthetic Be stars of the sample. The vertical axis shows the $\rm R - H\alpha$ colour, where higher values are brighter in $\rm H\alpha.$ Following \cite{Iqbal2013}, we calibrated the $\rm R - H\alpha$ colour of diskless main sequence stars to zero; this was done by shifting all the Be stars in the sample by +0.141 magnitudes in $\rm R - H\alpha$ such that the weighted average (by mass according to the initial mass function \citep{Salpeter1955}) of the stars with purely photospheric \ha\ profiles has an $\rm R - H\alpha$ colour of zero. The horizontal axis shows the $\rm V - I$ colour. We calibrated the sample stars in $\rm V - I$ according to the calibration standards of \cite{Ducati2001}; this was done by shifting all the Be stars in the sample by +0.346 magnitudes. After calibration, none of the photospheric stars in the sample differed from their associated calibration standards by more than 0.01 magnitudes in $\rm V - I$. The black line is a threshold line ($\rm R - H\,\alpha = 0.1(V-I) + 0.15$) above which any star is identified as a Be star candidate by the method of \cite{Iqbal2013}. In \cite{Iqbal2013}, both the slope and y-intercept of the threshold line are allowed to vary slightly depending on the star cluster being studied. Figure \ref{ikplot} uses the threshold line associated with the search for Be star candidates in NGC~330, which was chosen because the method of \cite{Iqbal2013} has recently been reassessed on this cluster \citep{Navarete2024}. 

The bottom panel of Figure~\ref{ikplot} shows the fraction of Be stars in the sample rejected as Be star candidates by the method of \cite{Iqbal2013}, namely those that fall below the threshold line, as a function of inclination. Again, while every star in the sample is a Be star whose \ha\ profile differs from a purely photospheric profile, $8,594$ out of $20,000$ ($\approx 43$\%) of the stars in the sample are rejected as Be star candidates, consistent with the method of \cite{Iqbal2013} selecting only strong \ha\ emitters. The shape of the distribution of rejected Be star candidates by inclination is relatively flat between $i=0^\circ$ and $40^\circ$, falls slightly between $50^\circ$ and $70^\circ$, and then rises steadily for $i\ge 80^\circ$. There is clear evidence of a selection bias against very high inclination objects: Be stars with an inclination of $90^\circ$ are the most likely to be rejected as candidates ($\approx 52\,\%$ are rejected) by the method of \cite{Iqbal2013} and are about $20\,\%$ more likely to be rejected than the average across all inclinations.     

\subsection{Milone et al.\ 2018}
\label{sec_mil18}

\cite{Milone2018} used observations from the Hubble Space Telescope's WFC3 to search 13 young clusters in the LMC and SMC for evidence of a ``split" main-sequence thought to represent stellar populations with differing rotation rates. \cite{Milone2018} used a photometric method to search for the presence of Be stars, which are known to be fast rotating main-sequence stars \citep{Rivinius2013}. The method of \cite{Milone2018} leverages the fact that stars with strong \ha\ emission will have smaller $\rm F656N - F814W$ magnitudes compared to stars without \ha\ emission at a given $F814W$. According to the method of \cite{Milone2018}, a Be star candidate is a star that lies to the left of a threshold line on a $F814W$ vs $\rm F656N - F814W$ colour-magnitude diagram. The threshold line was created by a shifting a fiducial line by -0.15 magnitudes (brighter in \ha); the fiducial line was created by fitting a cubic spline to the main-sequence stars of the cluster. As our sample of stars is synthetic, we recreated the fiducial line by fitting a cubic spline to the purely photospheric profiles only.

\begin{figure}
\includegraphics[width=0.48\textwidth ]{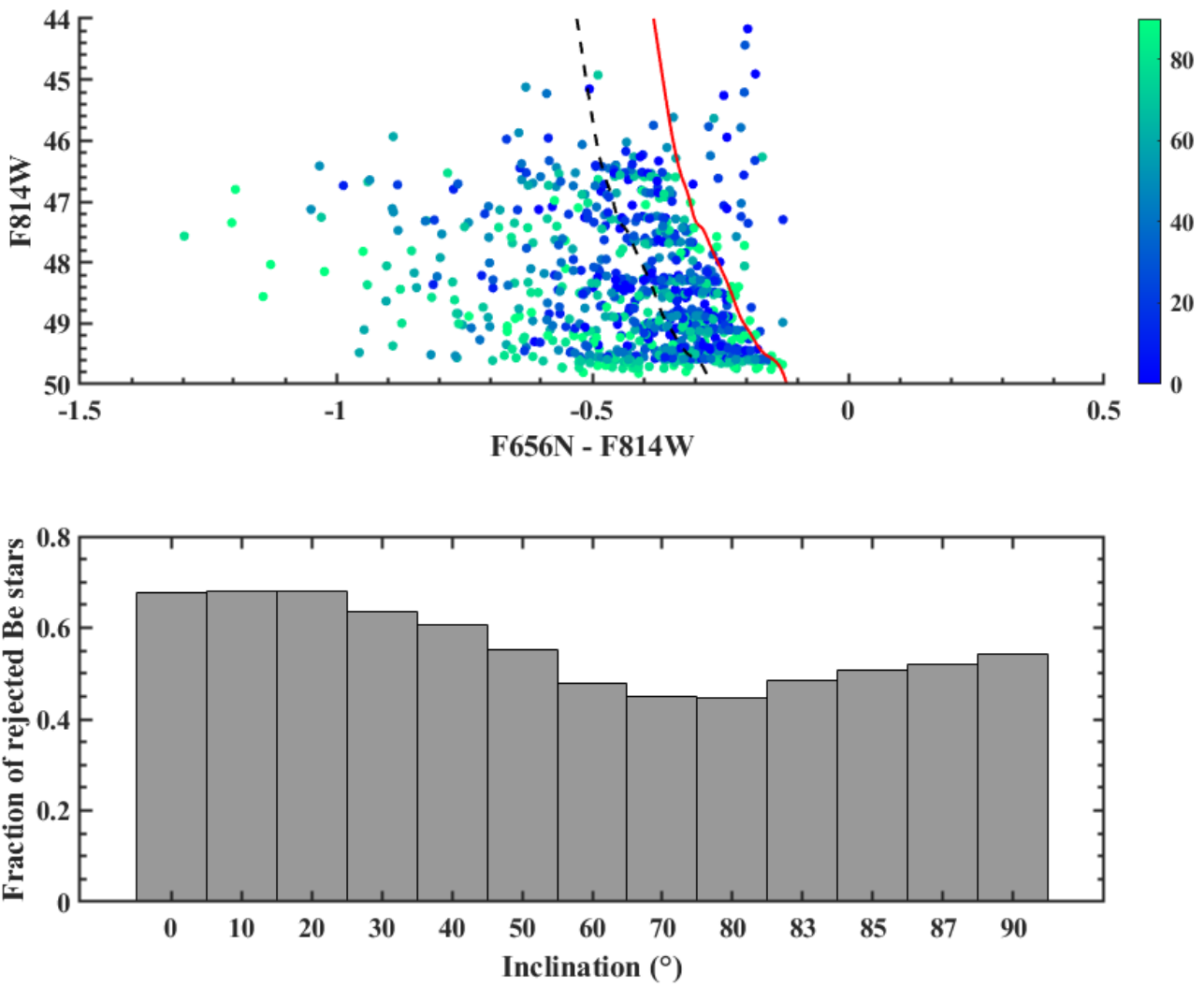}
\caption{Top: An $\rm F814W$ vs $\rm F656N-F814W$ colour-magnitude diagram used to identify Be star candidates based on the method of \cite{Milone2018}. Shown as circles are 1,000 randomly selected sample Be stars, with inclination coded according to the colourbar on the right. The red line is a cubic spline fit to the purely \ha\ photospheric profiles, and the dashed black line is a threshold line created by shifting the red line by -0.15 magnitudes (see text). According to the method of \citet{Milone2018}, stars left of this threshold line are Be star candidates. {\it Bottom}: A histogram of the fraction sample Be stars rejected as Be star candidates as a function of inclination. Note that low inclination Be stars are the most likely to be rejected, in contrast with Figure~\ref{ikplot}.}    
\label{Mil18plot}
\end{figure}

The top panel of Figure~\ref{Mil18plot} shows the method of \cite{Milone2018} applied to the Be stars of the sample. The vertical axis shows the $\rm F814W$ magnitude and the horizontal axis shows the $\rm F656N - F814W$ colour (lower values are brighter in $\rm H\alpha$). The red line is the fiducial line created by fitting a cubic spline to the purely photospheric profiles. The black dashed line is the threshold line created by shifting the fiducial line by -0.15 magnitudes in $\rm F656N - F814W$. As the method of \cite{Milone2018} ultimately depends only on the horizontal distance to a fiducial line derived from the purely photospheric profiles, it is unaffected by calibration shifts in $\rm F814W$. For this reason, the stars in Figure \ref{Mil18plot} have been left uncalibrated.

The bottom panel of Figure \ref{Mil18plot} shows the fraction of Be stars in the sample rejected as Be star candidates by the method of \cite{Milone2018} as a function of inclination. Again, although every star in the sample is a Be star by construction, 11,562 out of 20,000 stars ($\approx 58\,\%$) are rejected as Be star candidates. The shape of the distribution of rejected Be star candidates is flat and maximal between $i=0^\circ$ and $30^\circ$, then falls between $40^\circ$ and $80^\circ$ before rising for $i\ge80^\circ$. Thus the method of \cite{Milone2018} is most strongly biased against low inclination Be stars unlike the method of \cite{Iqbal2013}, which was most strongly biased against very high inclination Be stars. 

Finally, a two-sample Kolmogorov-Smirnoff (KS) test \citep{kolmogorov_1951} was used to compare the inclination distribution of rejected Be stars from the methods of \cite{Iqbal2013} and \cite{Milone2018} (that is, the bottom panels of Figures \ref{ikplot} and \ref{Mil18plot} multiplied by the number of stars in the sample at each inclination). The result rejects the null hypothesis that the two distributions come from the same underlying distribution with a a p-value of $0.031$.

\subsection{Discussion}

The results of Sections~\ref{sec_ik13} and \ref{sec_mil18} show that relatively similar photometric methods for identifying Be star candidates can produce somewhat different inclination biases, and there is likely no simple inclination bias that will apply to every photometric method. This section explores the causes of these differences in inclination bias, {\it i.e.\/} why the methods of \cite{Iqbal2013} and \cite{Milone2018} produce their maximum bias at opposite ends of the inclination range.     

The primary differences between the methods of \cite{Iqbal2013} and \cite{Milone2018} are the colours used ($\rm R - H\,\alpha$ vs $\rm F656N - F814W$), and the threshold line morphology (linear versus cubic spline of differing `slopes' through a secondary axis $\rm V - I$ vs $F814W$). A consequence of these differences is that the two methods reject different proportions of Be stars in the sample as Be star candidates ($43\,\%$ vs $58\,\%$).

\begin{figure}
\includegraphics[width=0.48\textwidth ]{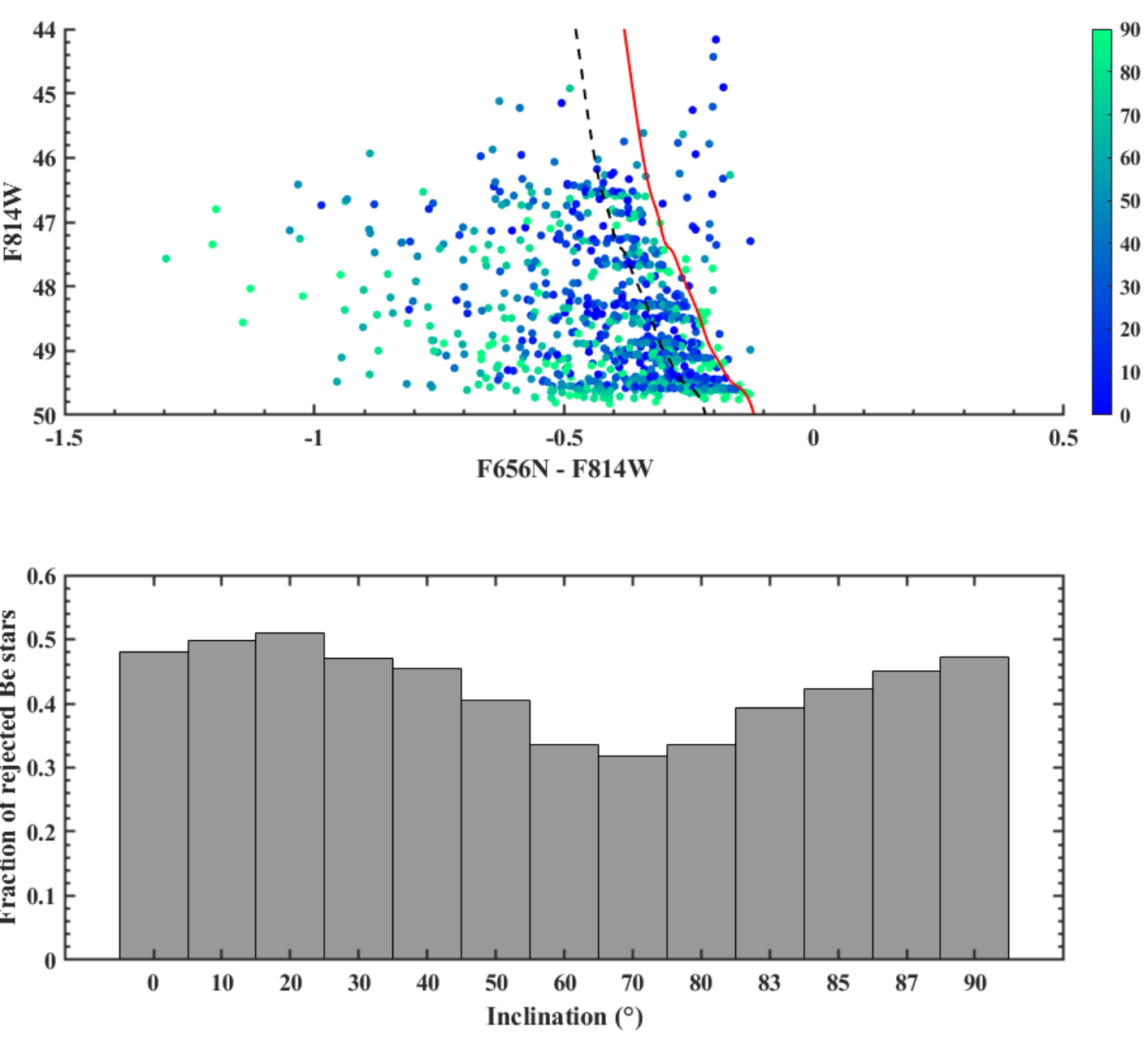}

\caption{{\it Top and Bottom}: The same as Figure~\ref{Mil18plot} except that the threshold was created by shifting the red line by $0.09715$ magnitudes, instead of $0.15$, to give the same number of rejected sample Be stars as the method of \cite{Iqbal2013} (see Section \ref{sec_ik13}).}
\label{Mil18plot09}
\end{figure}

Figure~\ref{Mil18plot09} examines the influence of the proportion of Be stars in the sample rejected as Be star candidates by recreating the method of \cite{Milone2018} with the modification that the threshold line is a shift of the fiducial line by -0.09715 magnitudes in $\rm F656N - F814W$ rather than the original -0.15 magnitudes. This change in the location of the threshold line results in the same number of rejected Be star candidates ($8,594 \approx 43\,\%$) as the method of \cite{Iqbal2013}. A two-sample KS test applied to the inclination distribution of rejected Be stars using both the method of \cite{Iqbal2013} and this modified method of \cite{Milone2018} accepts the null hypothesis that the two distributions come from the same underlying distribution with a p-value of $0.31$. Thus, the two photometric methods have inclination biases that are consistent with one another when the level of rejection is taken into account. However, the analysis is extended below to look at the impact of the two remaining differences between the methods of \cite{Iqbal2013} and \cite{Milone2018}: colour and threshold line morphology.

\begin{figure}
\includegraphics[width=0.48\textwidth ]{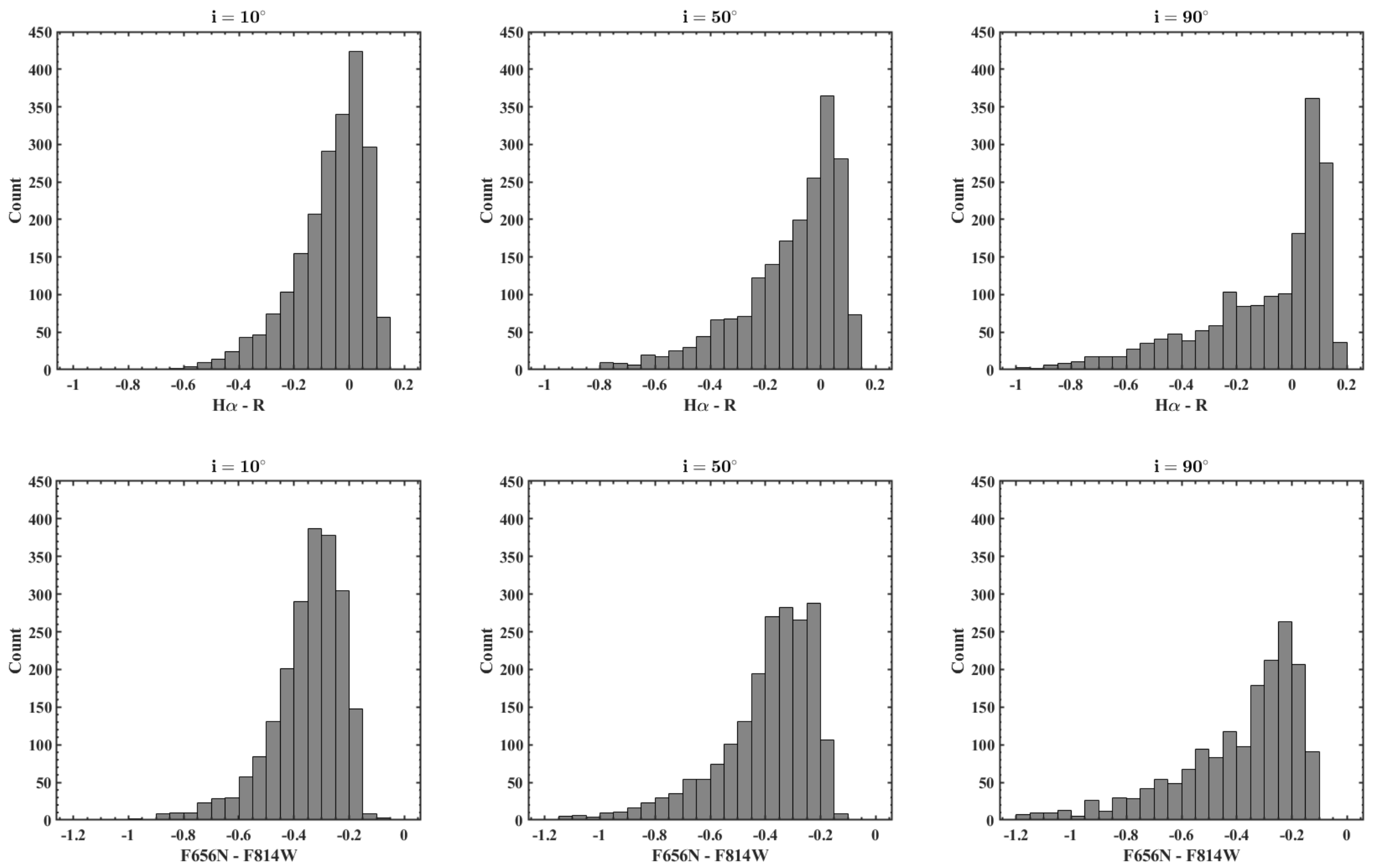}
\caption{Histograms in $\rm H\alpha - R$ (top row) and $\rm F656N - F814W$ (bottom row) for low ($i = 10^\circ$, left), medium ($i = 50^\circ$, middle), and very high ($i = 90^\circ$, right) inclination sample Be stars binned in colour increments of 0.05 magnitudes. Note that although the colour used in the method of \cite{Iqbal2013} is $\rm R - H\alpha$, it has been reversed in this figure to make comparisons with the $\rm F656N - F814W$ distributions easier.}    
\label{hist_phot_discuss}
\end{figure}

Moving on from the relationship between biases in inclination and the proportion of Be stars rejected as Be star candidates, we turn to the role of colour. Figure~\ref{hist_phot_discuss} is a panel plot of six histograms where the counts are over the synthetic Be stars in the sample with a given inclination and the bins are in colour increments of 0.05 magnitudes. Along the columns, each panel corresponds to low ($i = 10^\circ$, left), medium ($i = 50^\circ$, middle), and high ($i = 90^\circ$, right) inclinations. The top row shows the colour distribution of sample stars in $\rm H\alpha - R$, used by the method of \cite{Iqbal2013} (reversed so that the stars brightest in \ha\ are on the left side of the distribution to make a visual comparison with the bottom row easier). The bottom row shows the colour distribution of sample stars in $\rm F656N - F814W$, used by the method of \cite{Milone2018}. The purpose of these histograms is to remove the effects of threshold line morphology (namely the small positive slope in the method of \cite{Iqbal2013} and the curved fiducial line in the method of \cite{Milone2018}) to focus on the differences that arise due to the two methods different choices of colour.      

\begin{table}
\centering
\begin{tabular}{c c c c c c c c } 
\hline
$i$ & $\mu$ & $\sigma$ & Skew & Kurt & Q1 & Med & Q3 \\
\hline
 \multicolumn{8}{c}{$\rm H\,\alpha - R$} \\ 
\hline
$10^\circ$ & -0.07  & 0.13 & -1.14 & 4.31 & -0.14  & -0.04 & 0.03 \\ 

$50^\circ$ & -0.11 & 0.19 & -1.40 & 5.14 & -0.20 & -0.05 & 0.04 \\   
    
$90^\circ$ & -0.10 & 0.24 & -1.24 & 3.96 & -0.23  & 0.00 & 0.09 \\ 

\hline
\multicolumn{8}{c}{$\rm F656N - F814W$} \\ 
\hline
$10^\circ$ & -0.35  & 0.13 & -1.33 & 5.56  & -0.41  &  -0.33 & -0.26 \\ 

$50^\circ$ & -0.40 & 0.18 & -1.52 & 6.08 & -0.47 & -0.35 & -0.27 \\   
    
$90^\circ$ & -0.40 & 0.23 & -1.36 & 4.74 & -0.51 & -0.32 & -0.22 \\   \hline

\end{tabular}
\caption{Summary statistics for each of the six histograms shown in Figure \ref{hist_phot_discuss}. The rows are organized by inclination ($i$) in two sets of three; the top set of three rows corresponds to the $\rm H\,\alpha - R$ histograms (top row of Figure \ref{hist_phot_discuss}) and the bottom set of three rows corresponds to the $\rm F656N - F814W$ histograms (bottom row of Figure \ref{hist_phot_discuss}). The included summary statistics are the mean ($\mu$), the standard deviation ($\sigma$), the skewness (Skew), the kurtosis (Kurt), the first quartile (Q1), the median (Med), and the third quartile (Q3).}
\label{hist_phot_colour_stats}
\end{table}        

 Table~\ref{hist_phot_colour_stats} shows summary statistics for each of the six histograms in Figure~\ref{hist_phot_discuss}. The means of the $\rm H\alpha - R$ histograms are higher than those of their $\rm F656N - F814W$ counterparts at all inclinations, which is expected because Be stars are typically brighter at red wavelengths than in the near infrared. The standard deviations increase with inclination and are nearly identical between the two colours. The histograms of both colours are very negatively skewed at all inclinations as can be seen in the long left tails of the distributions in Figure~\ref{hist_phot_discuss}. This is corroborated by the mean being more negative than the median for each of the six histograms. Kurtosis values above three imply that each of the six panels contains more profiles in the wings of the distribution than would be expected if they were normally distributed; that the kurtosis values of the $\rm F656N - F814W$ histograms are higher than those of their $\rm H\alpha - R$ counterparts at all inclinations shows that they have more counts in the tails. Of the summary statistics listed in Table~\ref{hist_phot_colour_stats}, the inclination dependent quartiles are especially important for understanding the relationship between a photometric identification method and the resulting bias in inclination.    

\begin{figure}
\includegraphics[width=0.47\textwidth ]{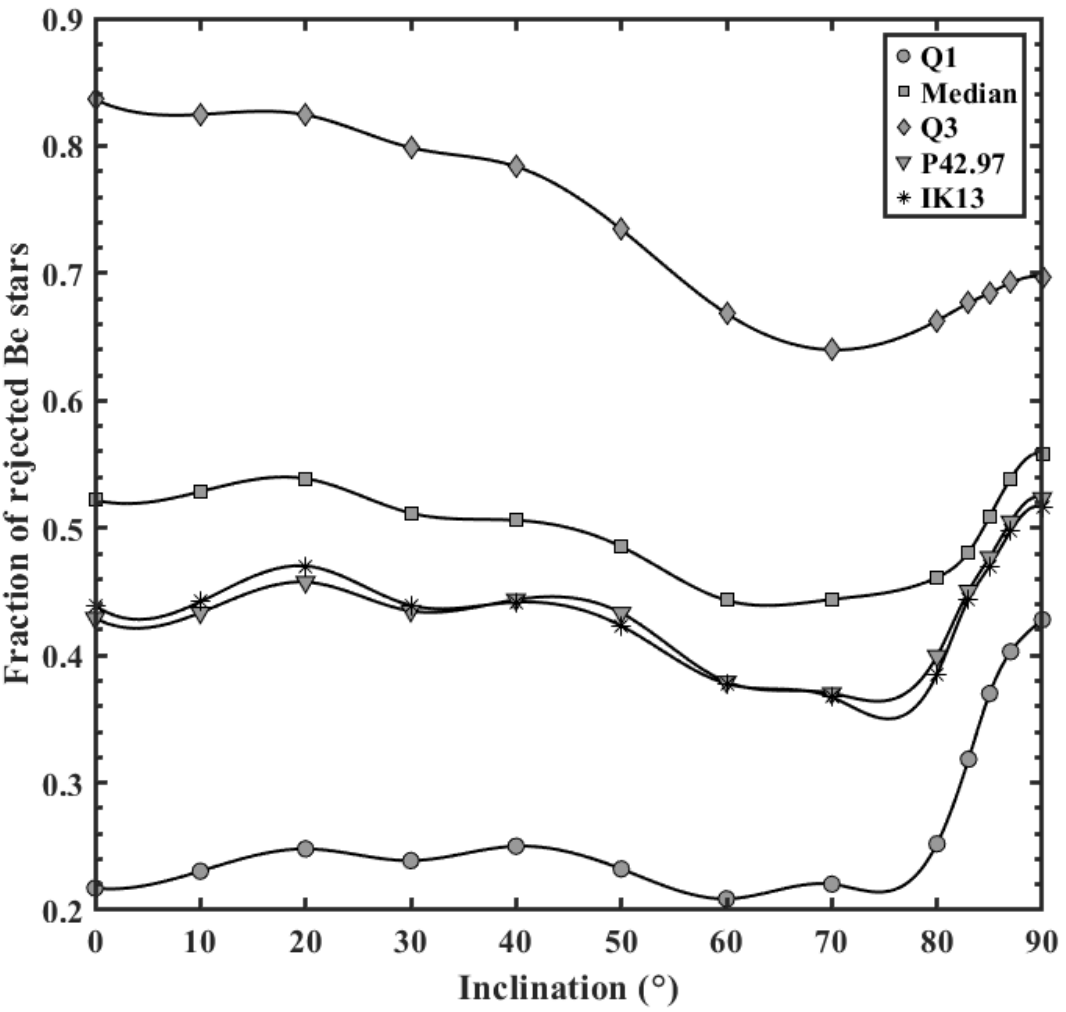}
\caption{A plot of the fraction of all the Be stars in the sample that were rejected as Be star candidates versus inclination for five different threshold lines in $\rm R - H\,\alpha$. The threshold lines correspond to the first quartile (circles), median (squares), third quartile (diamonds), 42.97th percentile (triangles), and the unmodified results using the method of \cite{Iqbal2013} (asterisks). The points have been fit with cubic splines to show the trends.}
\label{ik13_col_hists}
\end{figure}

Figure \ref{ik13_col_hists} shows the fraction of all the synthetic Be stars in the sample that were rejected as Be star candidates according to the method of \cite{Iqbal2013} with the modification of using four different quartiles in $\rm R - H\,\alpha$ as alternative threshold lines; the fraction of synthetic Be stars rejected using the unmodified method of \cite{Iqbal2013} is also included for comparison. These threshold lines occur at the:\\ 
\hspace*{0.3cm}$\bullet$\ First quartile (Q1) in $\rm R - H\,\alpha$ ($\rm R - H\,\alpha \approx 0.075$, shown as circles).\\
\hspace*{0.3cm}$\bullet$\ Median (Median) in $\rm R - H\,\alpha$ ($\rm R - H\,\alpha \approx 0.160$, shown as squares). \\
\hspace*{0.3cm}$\bullet$\ Third quartile (Q3) in $\rm R - H\,\alpha$ ($\rm R - H\,\alpha \approx 0.295$, shown as diamonds).\\
\hspace*{0.3cm}$\bullet$\ 42.97th percentile (P42.97) in $\rm R - H\,\alpha$ ($\rm R - H\,\alpha \approx 0.133$, shown as triangles).\\
\hspace*{0.3cm}$\bullet$\ Unmodified location of the method of \cite{Iqbal2013} (IK13) ($\rm R - H\,\alpha = 0.1(V-I) + 0.15$, shown as asterisks).\\
The threshold line corresponding to the 42.97th percentile in $\rm R - H\,\alpha$ was chosen because this results in the same number of rejected Be stars (8,594) as the unmodified method of \cite{Iqbal2013} (see Section~\ref{sec_ik13}), allowing the relationship between line profile morphology and inclination bias to be explored when the fraction of rejected Be stars and the colours are held constant but the slope is set to zero. Finally, the unmodified method of \cite{Iqbal2013} contains identical values to the bottom panel of Figure \ref{ikplot}). 

Threshold line morphology turns out to be relatively unimportant to the method of \cite{Iqbal2013} (the small slope through V-I was likely chosen to avoid the inclusion of cool red giant branch objects); a two-sample KS test between the P42.97 (triangles) and IK13 (asterisks) distributions of Figure~\ref{ik13_col_hists} (multiplied by the number of stars in the sample at each inclination) strongly accepts the null hypothesis that the two distributions come from the same underlying distribution with a p-value of 0.97. While the effect is small, the threshold line used in the method of \cite{Iqbal2013} rejects slightly more low inclination objects and slightly less high inclination objects than the threshold line at the 42.97th percentile in $\rm R - H\,\alpha$.  

At high inclinations, \ha\ line profiles can feature deep absorption troughs that sometimes result in a Be star having higher $\rm R - H\,\alpha$ magnitudes than a purely photospheric star. This effect is especially prominent above $i = 80^\circ$ and manifests in the Q1 distribution of Figure \ref{ik13_col_hists} being the most biased against high inclinations because these high $\rm R - H\,\alpha$ magnitude Be stars represent a greater proportion of the total number of rejected stars. The effect can also be seen in Figure~\ref{hist_phot_discuss} by noting that a far higher fraction of $i = 90^\circ$ Be stars are dim in $\rm H\alpha$ (i.e., have higher counts between 0.1 and 0.2 on the top row) compared to their $i = 10^\circ$ or $i = 50^\circ$ counterparts. 
 
 As we move to threshold lines that reject a higher fraction of the total number of Be stars in the sample as Be star candidates, the high $\rm R - H\,\alpha$ magnitude Be stars that are disproportionately present at high inclinations become a smaller fraction of the total number of rejected Be stars; this, coupled with low inclination Be stars having fewer objects in the tail of their $\rm R - H\,\alpha$ distribution results in the bias becoming strongest against low inclination Be stars in the Q3 distribution of Figure~\ref{ik13_col_hists}. 
 
 Another notable feature of Figure~\ref{ik13_col_hists} is the dip in rejected Be stars that reaches a minimum between $60^\circ$ and $70^\circ$. This can be understood in terms of the two previously mentioned effects: unlike $i = 90^\circ$ Be stars, moderate inclination Be stars do not feature an excess of high $\rm R - H\,\alpha$ objects resulting from \ha\ line profiles that feature deep absorption troughs and while there are less moderate inclination Be stars in the tail of the \ha\  distribution compared to $i = 90^\circ$ Be stars, there are significantly more than for low inclination Be stars. The combination of these two effects results in the dip in rejected Be stars seen at moderate inclinations. The distribution of rejected Be stars obtained by the method of \cite{Iqbal2013} is intermediate between the Q1 and Median distributions.       

\begin{figure}
\includegraphics[width=0.47\textwidth ]{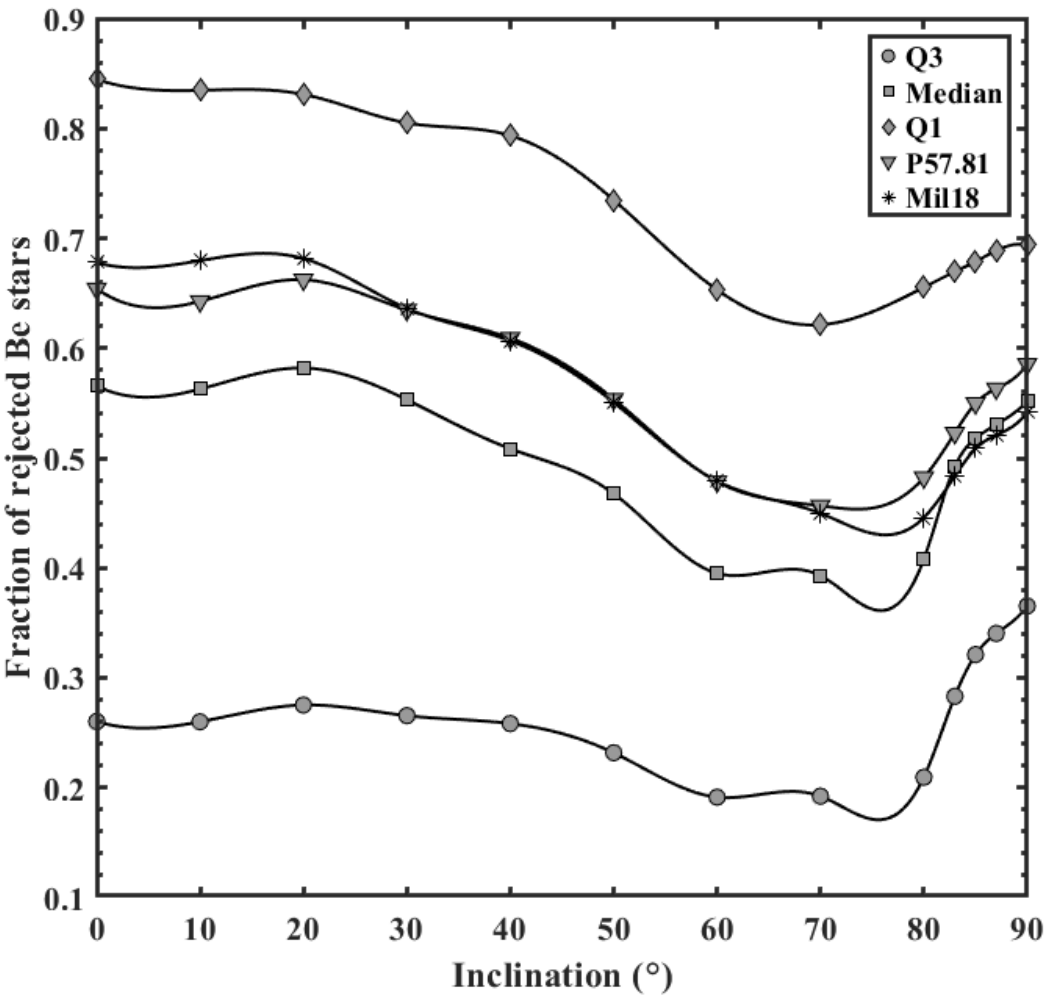}
\caption{A plot of the fraction of all the Be stars in the sample that were rejected as Be star candidates versus inclination for five different threshold lines in $\rm F656N - F814W$. The threshold lines correspond to the first quartile (diamonds), median (squares), third quartile (circles), 57.81th percentile (triangles), and the unmodified results using the method of \cite{Milone2018} (asterisks). The points have been fit with cubic splines to show the trends. Note that compared to Figure \ref{ik13_col_hists}, the symbol used for the first and third quartiles have been swapped to aid comparison.}
\label{mil18_col_hists}
\end{figure}

Figure~\ref{mil18_col_hists} shows a plot of the fraction of all the synthetic Be stars in the sample that were rejected as Be star candidates according to the method of \cite{Milone2018} with the modification of using four different quantiles in $\rm F656N - F814W$ as alternative threshold lines; the fraction of synthetic Be stars rejected using the unmodified method of \cite{Milone2018} is also included for comparison. The threshold lines occur at the:\\ 
\hspace*{0.3cm}$\bullet$\ Third quartile (Q3) in $\rm F656N - F814W$ ($\rm F656N - F814W \approx -0.261$, shown as circles).\\
\hspace*{0.3cm}$\bullet$\ Median (Median) in $\rm F656N - F814W$ ($\rm F656N - F814W \approx -0.344$, shown as squares).\\ 
\hspace*{0.3cm}$\bullet$\ First quartile (Q1) in $\rm F656N - F814W$ ($\rm F656N - F814W \approx -0.462$, shown as diamonds).\\
\hspace*{0.3cm}$\bullet$\ 57.81th percentile (P57.81) in $\rm F656N - F814W$ ($\rm F656N - F814W \approx -0.373$, shown as triangles).\\
\hspace*{0.3cm}$\bullet$\ Unmodified location of the method of \cite{Milone2018} (Mil18) (see Figure \ref{Mil18plot}, shown as asterisks).\\
The positions of the first and third quartiles have been swapped to make comparisons between the two sets of histograms easier ({\it i.e.\/} the first quartile in $\rm R - H\,\alpha$ rejects the same number of profiles, 5,000 out of 20,000 or 25\%, as the third quartile in $\rm F656N - F814W$). The threshold line corresponding to the 57.81th percentile in $\rm F656N - F814W$ was chosen because it results in the same number of rejected Be stars (11,562) as the unmodified method of \cite{Milone2018} (see Section~\ref{sec_mil18}), allowing us to explore the relationship between line profile morphology and inclination bias when the fraction of rejected Be stars and the colour are held constant. Finally, the unmodified method of \cite{Milone2018} (containing identical values to the bottom panel of Figure \ref{Mil18plot}) is included for comparison. 

While more important than for the method of \cite{Iqbal2013}, threshold line morphology remains relatively unimportant to the method of \cite{Milone2018}; a two-sample KS test between the P57.81 (blue) and Mil18 (turqoise) distributions of Figure~\ref{mil18_col_hists} (multiplied by the number of stars in the sample at each inclination) accepts the null hypothesis that the two distributions come from the same underlying distribution with a p-value of 0.78. The threshold line used in the method of \cite{Milone2018} rejects more low inclination objects and less high inclination objects than the threshold line at the 57.81th percentile in $\rm F656N - F814W$ (an effect that was also observed with the method of \cite{Iqbal2013}; see Figure \ref{ik13_col_hists}).  

 As previously noted, \ha\ line profiles can feature deep absorption troughs at high inclinations that sometimes result in a Be star having higher $\rm F656N - F814W$ magnitudes (dimmer in $\rm H\alpha$) than a purely photospheric star. This effect manifests in the Q3 distribution of Figure~\ref{mil18_col_hists} being the most biased against $i = 90^\circ$ Be stars and can also be seen in Figure \ref{hist_phot_discuss} by noting that a far higher fraction of $i = 90^\circ$ Be stars are dim in \ha\ (with higher counts between -0.2 and -0.1 on the bottom row) compared to their $i = 10^\circ$ or $i = 50^\circ$ counterparts. Bias against low inclinations as the proportion of rejected Be stars increases is driven by the relative lack of Be stars in the (bright in \ha) tail of their $\rm F656N - F814W$ distribution (see the bottom row of Figure~\ref{hist_phot_discuss}). This bias against low inclination angles is seen most clearly in the Q1 distribution of Figure~\ref{mil18_col_hists}. There is a dip in the fraction of rejected Be stars that reaches a minimum between between $60^\circ$ and $70^\circ$. This occurs because, unlike high inclination Be stars, these moderate inclination Be stars do not contain high $\rm F656N - F814W$ values (dim in \ha) resulting from \ha\ line profiles that feature deep absorption troughs while also containing more Be stars in the bright in \ha\ tail of their $\rm F656N - F814W$ distribution compared to lower inclination Be stars. The distribution of rejected Be stars obtained by the method of \cite{Milone2018} is intermediate between the Median and Q1 distributions. 

 Comparing the biases produced by photometric methods based on $\rm R - H\,\alpha$ (Figure~\ref{ik13_col_hists}) with those based on $\rm F656N - F814W$ (Figure \ref{mil18_col_hists}), the former is more biased against very high inclination Be stars when the proportion of Be stars in the sample rejected as Be star candidates is low (compare Q1 in Figure~\ref{ik13_col_hists} to Q3 in Figure~\ref{mil18_col_hists}). Conversely, there is a stronger bias produced by photometric methods based on $\rm F656N - F814W$ than by those based on $\rm R - H\,\alpha$ against low inclinations when the proportion of rejected Be stars is high (compare Q3 in Figure~\ref{ik13_col_hists} to Q1 in Figure~\ref{mil18_col_hists}). Both effects occur because the $\rm F656N - F814W$ distribution has a heavier, bright in \ha, tail than the $\rm R - H\,\alpha$ distribution (compare the top and bottom rows of Figure \ref{hist_phot_discuss} and the kurtosis column of Table \ref{hist_phot_colour_stats}).

\begin{figure}
\includegraphics[width=0.47\textwidth ]{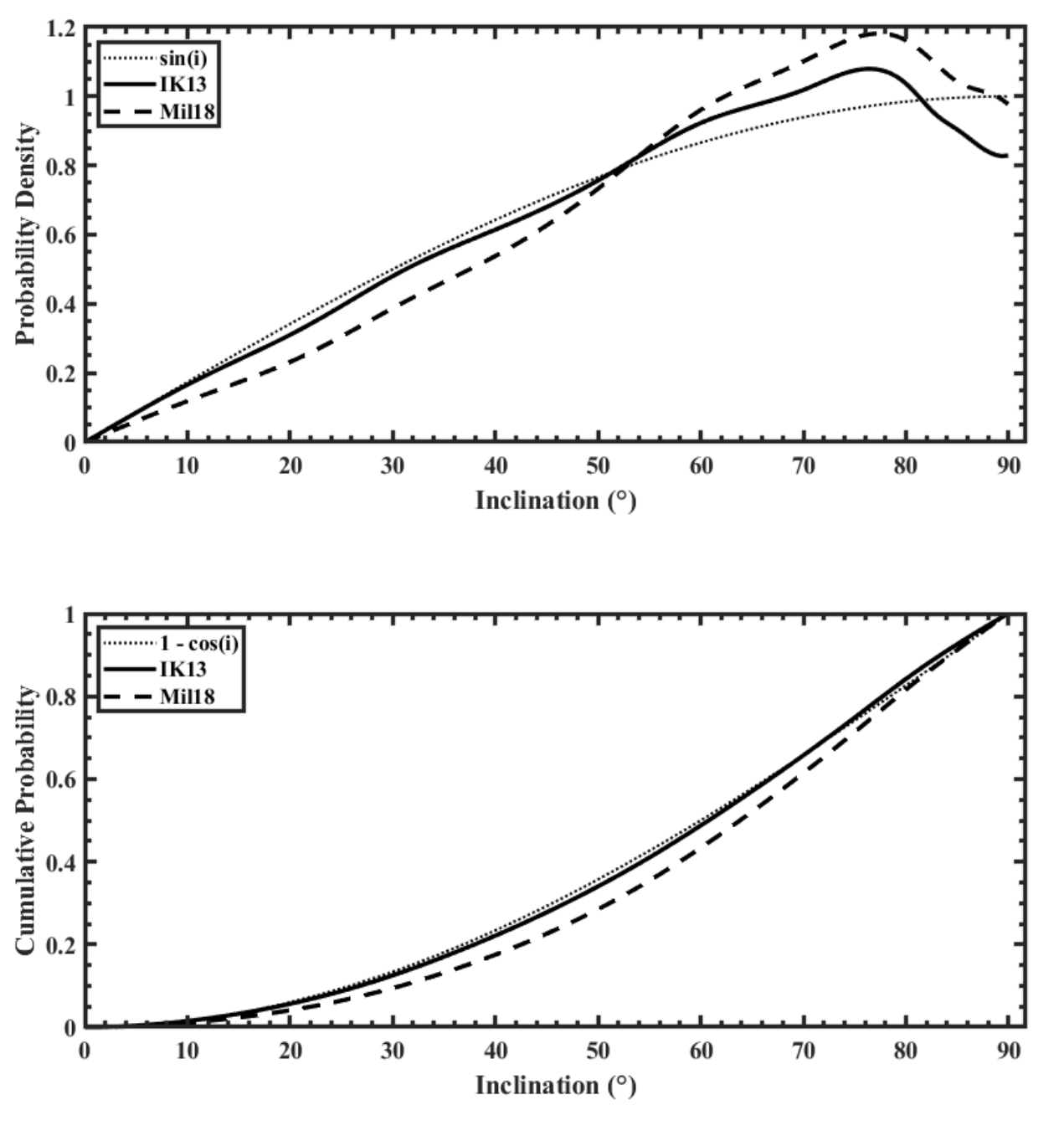}
\caption{The inclination biases of the methods of \cite{Iqbal2013} and \cite{Milone2018} determined on the sample of synthetic Be star spectra expressed as both probability density functions (top panel) and cumulative distribution functions (bottom panel). In both panels, the inclination bias of the method of \cite{Iqbal2013} (IK13) is shown as a solid line, the inclination bias of the method of \cite{Milone2018} (Mil18) is shown as a dashed line, and the $\sin{i}$ probability distribution is shown as a dotted line for comparison.}
\label{pdf_phot}
\end{figure}

Figure~\ref{pdf_phot} shows the observational inclination biases of the methods of both \cite{Iqbal2013} and \cite{Milone2018} expressed as probability density functions (top panel) and cumulative distribution functions (bottom panel). These pdfs were created in the same manner as Figure~\ref{pdf_spectroscopic} using Eq.~(\ref{eq:ipdf}). Compared to the $\sin{i}$ distribution, the methods of both \cite{Iqbal2013} and \cite{Milone2018} are biased against low ($i < 50^\circ$) inclination Be stars, the latter significantly more so than the former. Both methods result in an excess of Be stars as Be star candidates at moderate inclinations ($50^\circ < i < 80^\circ$), and again this bias is stronger for the method of \cite{Milone2018}. At very high inclinations ($ i \approx 90^\circ$) the inclination biases of the two methods diverge in an interesting way: the method of \cite{Milone2018} is relatively unbiased whereas the method of \cite{Iqbal2013} is significantly biased against these high inclination Be stars. Overall, the method of \cite{Iqbal2013} is less biased in inclination on our sample than the method of \cite{Milone2018}, except near $i=90^\circ$.

\section{Conclusions and Future Work}
\label{conclusions}

The spectroscopic method of \cite{Hou2016} and the photometric methods of \cite{Iqbal2013} and \cite{Milone2018} for identifying Be star candidates were tested for bias with respect to system inclination using a sample of 20,000 synthetic Be stars. The spectroscopic method showed substantial bias against high inclinations ($i > 80^\circ$). The photometric methods were both biased against low inclinations and one of the photometric methods was also biased against inclinations above $80^\circ$, resulting in a surplus in the detection rate of moderate inclination ($ 50^\circ < i < 80^\circ$) Be star candidates. Each of the three methods for identifying Be star candidates considered in this work has a unique bias in inclination and these biases will have to be accounted for in order to use Be stars as a probe for correlated spin-axes in young open clusters or to address the apparent lack of high inclination Be stars. As spectroscopic surveys continue to grow in size and machine learning approaches to stellar categorization become increasingly prevalent \citep[e.g.,][]{Vioque2020,Wang2022}, accounting for the inclination biases in the samples of Be stars produced by these approaches will become necessary to properly interpret the resulting statistical analyses.

An interesting avenue for future work is determining potential inclination biases of methods based on photometric time-series. Methods for determining Be star candidates based on photometric time-series approaches have re-emerged as an area of active research \citep[e.g.,][]{Ortiz2017,Vioque2020,Granada2023}, owing to Gaia photometry, after achieving initial success in the early 2000s \citep[e.g.,][]{Keller2002,Mennickent2002}. Additionally, some studies incorporate both spectroscopic and photometric methods to detect Be star candidates \citep[e.g., Figure~8 of][]{Marino_2018}, which could be a powerful way to compare and calibrate approaches based on the two classes of methods. We note that in some cases, such as the methods \citet{Hou2016} and \cite{Milone2018}, the inclination biases are complementary and could potentially be used together to reduce the inclination bias compared to using either method alone. With sufficiently many stars, acceptance as a Be star candidate using a spectroscopic method could be further examined as a function of photometrically derived properties such as the position on a colour-magnitude diagram or the distance from a threshold line.

It could be possible to use the simulation-based approach of this work to provide insight into why Be stars in different clusters exhibit different F656N-F814W color distributions \cite[e.g.\ Figure~18 of][]{Milone2018}. New simulated samples, based on the current approach, could be computed to represent young, open clusters with varying distributions of metallicity \cite[including the effect on the Be star disks; see][]{Ahmed2012}, rotational velocity, and cluster age. The resulting F656N-F814W colour distributions of the simulated clusters could then be compared with observed SMC data to determine if the observed distributions are successfully recovered. In agreement with Figure~18 of \cite{Milone2018}, \cite{Radley2025} noted that the colour excess of Be stars compared to main sequence B-type stars tends to increase towards earlier subtypes which may help to explain these colour distributions.

Finally, in the synthetic sample analyzed in this work, a significant fraction of the model Be stars are rejected as Be star candidates (between $\approx 24\%$ and $\approx 58\%$, depending on the method used to determine Be star candidacy). As the fraction of detected Be stars can exceed 50 percent near the main sequence turnoff in LMC/SMC clusters \citep{Milone2018,Bodensteiner2020}, one might wonder if it is possible that all such bright main-sequence B stars are Be stars. This is an interesting question, but it is hard to assess from the model disks upon which our conclusions are based. While we attempted to exclude model disk parameters ($\rho_0$,n,$R_d$) that resulted in weakly-emitting disks that would never be detected as Be stars (via comparison with the underlying photospheric profile), we do not know if all models passing this test are physically realized and, if so, with what frequency. All we can conclude is that if such disks do occur, the candidate selection methods investigated here would miss a large fraction of these Be stars; therefore, one should be cautious about interpreting observational determinations of Be star fractions using candidates selected using these methods.

\begin{acknowledgments}
B.\ D.\ Lailey acknowledges support from the University of Western Ontario's physics and astronomy department. T.\ A.\ A.\ Sigut acknowledges support in the form of a Discovery Grant from the Natural Sciences and Engineering Council of Canada. The authors thank the anonymous referee for the thoughtful feedback. 
\end{acknowledgments}

\software{Astropy \citep{astropy:2013, astropy:2018, astropy:2022}, \texttt{Bedisk} \citep{Sigut2007}, \texttt{Beray} \citep{Sigut2018}, Violin Plots for Matlab (Bechtold 2016) }

\bibliography{main.bib}
\bibliographystyle{aasjournal}

\end{document}